\documentclass[aps,reprint,twocolumn,pra,groupedaddress,floatfix]{revtex4-2}
\usepackage{amsmath,amssymb,amsfonts}
\usepackage{mathrsfs}
\usepackage{stackengine}
\usepackage{soul}
\usepackage{graphicx}
\usepackage{color}
\usepackage{multirow}
\begin{document}
	\title{Unique multistable states in periodic structures with saturable nonlinearity}
	\author{S. Vignesh Raja$^{\ddagger}$}
	\author{A. Govindarajan$^{\dagger}$}
	\email[Corresponding author: ]{govin.nld@gmail.com}
	\author{M. Lakshmanan$^{\dagger}$}
	\affiliation{$^{\ddagger}$Department of Physics, Pondicherry University, Puducherry, 605014, India}
	
	\affiliation{$^{\dagger}$Department of Nonlinear Dynamics, School of Physics, Bharathidasan University, Tiruchirappalli - 620 024, India}
	\begin{abstract}
		
		We report that conventional saturable periodic structures, in sharp contrast to the conventional systems with different nonlinearities which exhibit the typical S- shaped  optical bi- and multi-stable states,  reveal some unusual and unique nonlinear dynamics. These include the onset of  ramp-like optical bistability (OB) and optical multistability (OM) curves which further transit into mixed OM states combining both ramp-like states followed by the S-shaped multistable curves. We also extend this study to another domain of physics, namely parity-time ($\mathcal{PT}$)- symmetry, by including equal amount of gain and loss into the system which then establishes additional degree of freedom by enabling the investigation into additional two domains which are the unbroken and broken $\mathcal{PT}$- symmetric regimes. Although these bi- and multi-stable states are unusual and unique, when the frequency detuning is introduced, the revival of S-shaped stable states is possible but only in the presence of unbroken $\mathcal{PT}$- symmetry. Conversely, the broken $\mathcal{PT}$- symmetry which usually generates ramp-like multistable states, gives rise to the birth of novel multistable states with a vortex like envelope, (the curve that features simultaneous increase in the critical switch-up and switch-down powers with an increase in the input power) causing a novel structure which has not been reported in the existing literature of different physical systems manifesting multi-stable states.
	\end{abstract}
	\maketitle
	\section{Introduction}
	\label{Sec:1}
	In optical fibers, nonlinearities (NLs) arise from the third-order susceptibility
	$\chi^{(3)}$, which is responsible for phenomena such as third-harmonic generation,
	four-wave mixing, and nonlinear refraction. Among these,  the third-harmonic generation and four-wave mixing that dealt with the generation of new frequencies are generally hard to realize in optical fibers unless phase matching is achieved. Thus, most of the nonlinear effects in optical fibers mainly stem from the nonlinear refraction, where the refractive index depends on the light intensity \cite{agrawal2000nonlinear,agrawal2001applications}. The induced nonlinearity depends on the laser power and the nature of the material. For instance, an ordinary silica fiber features a low nonlinear coefficient value and induces third-order cubic nonlinearity alone, and higher-order nonlinearities are uncommon in them \cite{agrawal2001applications}. Experimental measurements show that the fifth and seventh-order nonlinear coefficients have a pivotal role in controlling the nonlinear phase shift induced by the high-intensity laser in chalcogenide fibers \cite{harbold2002highly,chen2006measurement}. The variations in the doping concentration along the fiber length create inhomogeneous nonlinear profiles \cite{nobrega1998optimum,nobrega2000multistable}.  
	
	Sulfide- \cite{hall1989nonlinear,olbright1986interferometric}, and heavy-metal-doped oxide \cite{kang1995femtosecond} glasses exhibit nonlinear saturation response. These materials possess a faster nonlinear and slower thermal response than silica fibers \cite{yao1985ultrafast}. In such cases, cubic nonlinearity may not accurately
	characterize the induced refractive index variations \cite{langbein1985nonlinear} at higher input intensities ($P_0$) \cite{da1995dynamics}. In reality, the nonlinear response of such materials cannot increase infinitely, and it saturates beyond a value of input intensity known as critical intensity. In other words, there is a maximum limit for the intensity-dependent refractive index change. Beyond this limit, the variations in the intensity-dependent refractive index cease \cite{herrmann1991propagation}. The intensity at which the nonlinear response saturates varies for different materials \cite{abou2011impact}. For instance, the effect of saturable nonlinearity (SNL) comes into play at moderately high intensities in $CdS_{x}Se_{1-x}$ semiconductor-doped glass fiber \cite{gatz1991soliton}.

	Periodic perturbations in the refractive index of the fiber in the form of Bragg gratings lead to the reflection of a band of input optical signals. The range of wavelengths reflected by the fiber Bragg grating (FBG) is called the photonic bandgap (PBG) or stopband \cite{erdogan1997fiber}. An association between PBG and intensity-dependent refractive index promotes the study of several nonlinear effects that include all-optical switching.  The anticipation for alternative solutions to control light with light has accelerated substantial research growth in nonlinear FBGs. Investigations on the steady-state switching dynamics of FBGs mainly target a reduction in the switching intensities \cite{agrawal2001applications}. Switching in an FBG is described by the OB or OM phenomenon \cite{winful1979theory}. As the name suggests, the transmission characteristics of nonlinear FBGs present two or more output states for a given input power. For studying OB in FBGs, researchers employed numerous materials with a variety of nonlinearities  \cite{karimi2012all,lee2003nonlinear,yosia2007double}.

	The primary theme of the article is the study of nonreciprocal switching dynamics in a grating structure in the presence of saturable nonlinearity (SNL). For the steering dynamics to be nonreciprocal, the nonlinear response of the device must be direction-dependent.  In other words, the input-output characteristic curves pertaining to the different light launching directions (left and right) should be distinguishable.  Acquiring asymmetric switching response requires the construction of geometrically
	asymmetric system (the classic approach) [see \cite{caloz18} and
	references therein]  or the incorporation of parity and time ($\mathcal{PT}$) symmetry in the form of gain and loss \cite{el2007theory, regensburger2012parity, kottos2010optical,feng2017non,ruter2010observation} (the modernistic approach) into the traditional FBG structures \cite{raja2019multifaceted,PhysRevA.100.053806,sudhakar2022low,komissarova2019pt}. Nonreciprocal mode interaction occurs when the periodic perturbation is a complex function of the form $n_0+n_{1R} cos( 2 \pi z/\Lambda) + i sin ( 2 \pi z/\Lambda)$. Physical realization of a parity and time symmetric-FBG (PTFBG) requires gain and loss regions to be placed next to each other in one period $\Lambda$ (unit cell) and periodically repeating the unit cell for the entire device length ($L$). The cosine and sine terms in the expansion of this complex exponential function signify the modulation in the real ($n_{1R}$) and imaginary ($n_{1I}$) parts of the complex refractive index profile [$n(z)$]. Such an arrangement ensures that the system obeys the $\mathcal{PT}$-symmetric condition $n(z)=n^*(-z)$ \cite{phang2015versatile,phang2014impact,phang2015versatile,govindarajan2018tailoring,govindarajan2019nonlinear}. 
	
	In the context of FBGs, Poladian conceptualized the complex periodic perturbation of the refractive index profile \cite{poladian1996resonance}, and Kulishov \emph{et al.} carried out the first systematic study on the linear response of the device \cite{kulishov2005nonreciprocal}. The groundwork laid by Kulishov \emph{et al.} paved the way for non-Hermitian Physicists to accomplish significant discoveries in linear PTFBGs \cite{raja2020phase,raja2020phase,raja2021n}.  Lin \emph{et al.} reiterated the same study done by Kulishov \emph{et al.} \cite{kulishov2005nonreciprocal} and pointed out that operating the nonlinear PTFBG at the unitary transmission point leads to the loss of bistable or multistable switching behavior \cite{lin2011unidirectional}. Several years ago, researchers thought that the nonlinear FBG switches function in the unbroken $\mathcal{PT}$-symmetric regime alone \cite{1555-6611-25-1-015102}. Due to the limited range of operation, the field of all-optical switching in nonlinear PTFBGs did not receive significant research interest in the past. 
	
	The literature presents a wide range of options to optimize the power required for switching like frequency detuning \cite{winful1979theory},  incorporating phase-shift region in the middle of FBGs (PSFBGs) \cite{radic1994optical}, introducing inhomogeneous nonlinear profiles \cite{sudhakar2022inhomogeneous}, higher-order nonlinearities \cite{raja2019multifaceted}, chirping in conventional and $\mathcal{PT}$-symmetric chirped FBGs \cite{maywar1998effect, PhysRevA.100.053806}, four-wave mixing \cite{sudhakar2022low} and by modifying the signal parameters like pulse width and shape in the time-domain  \cite{lee2003nonlinear}. Contemporary works on the steady-state switching dynamics of PTFBGs validate that the gain and loss parameter ($g$) impacts in reducing the switch-up and down intensities, provided that its magnitude is closer to the value of the coupling parameter ($\kappa$) \cite{raja2019multifaceted, PhysRevA.100.053806,sudhakar2022low}.  In Ref. \cite{raja2019multifaceted}, the existence of ramp-like OB (OM) curves confirms that the broken regime is not an instability domain.  The concept of launching light from the rear end serves as a new route for realizing low-power switches. \cite{litchinitser2007optical,venugopal2011,komissarova2019pt,raja2019multifaceted,PhysRevA.100.053806}.

Researchers have investigated different types of nonlinear FBGs and PTFBGs in the past from a switching viewpoint without including SNL. The scientific contributions that deal with the impact of SNL on the dynamics exhibited by non-periodic structures are many. Nevertheless, there seem to exist no works dealing with switching dynamics exhibited by nonlinear periodic structures with SNL. Therefore, we present the mathematical function that describes SNL in periodic structures in Sec. \ref{Sec:2}. Furthermore, this section also deals with the derivation of the first-order differential equations or coupled mode equations of the present system in detail.  The conventional model was used in the literature to study soliton dynamics in periodic structures \cite{malomed2005coupled}. Since the seminal proposal by Merhasin \emph{et al.} \cite{merhasin2007solitons}, the literature does not find any systematic research on FBGs with SNL from switching or any other application perspectives. For the first time, we investigate the switching dynamics shown by a conventional FBG with SNL. We also wish to know whether the SNL parameter and $\mathcal{PT}$-symmetry  alters the characteristics of the hysteresis curves and the switching intensities.

	\section{Theoretical framework}
	\label{Sec:2}
	The refractive index distribution [$n(z)$]  of a PTFBG that includes the SNL effect reads as
	\begin{gather}
		\nonumber n(z)=n_0+n_{1R} cos(2 \pi z/\Lambda)+in_{1I} sin(2 \pi z/\Lambda)\\-{n_2}f(|E|^2).
		\label{Eq:1}
	\end{gather}
	Squaring Eq. (\ref{Eq:1}) and neglecting higher-order terms in $n_{1I}$, $n_{1R}$, and $n_2$ results in 
	\begin{gather}
		\nonumber n^2(z)=n_0^2+2 n_o n_{1R} cos(2 \pi z/\Lambda)+2i n_0 n_{1I} sin(2 \pi z/\Lambda)\\+2 n_0 {n_2}f(|E|^2).
		\label{Eq:2}
	\end{gather}
	The function $f(|E|^2)=\cfrac{1}{1+|E|^2}$ better describes the SNL in periodic structures and has a nonlinear term in the denominator \cite{malomed2005coupled,merhasin2007solitons,yulin2008discrete,melvin2006radiationless}. The proposed model is analogous to the discrete version of the Vinetskii–Kukhtarev equation that accounts for the SNL in 1-dimensional optical lattices and waveguide arrays \cite{hadvzievski2004power,vicencio2006discrete}. In Eq. (\ref{Eq:2}), $n_0$ and $n_2$ represent the constant refractive index and nonlinear coefficient of the FBG, respectively. The refractive index perturbations ($n_{1R}$, $n_{1I}$ and $n_2$) are small compared to the core refractive index ($n_0$). At a given operating wavelength ($\lambda$), the coupling parameter ($\kappa$) dictates the amount of coupling  between the counter-propagating fields, and its mathematical representation reads $\kappa=\cfrac{\pi n_{1R}}{\lambda}$. The balanced gain and loss levels ($g$) supplied to achieve $\mathcal{PT}$-symmetry depend on the modulation of $n_{1I}$, and the relation between these two parameters reads as  $g=\cfrac{\pi n_{1I}}{\lambda}$. The saturable nonlinearity parameter ($S$) is mathematically related to the nonlinear coefficient of the material ($n_2$) via the mathematical expression, $S =\cfrac{2 \pi n_2}{\lambda}$ \cite{raja2019multifaceted,PhysRevA.100.053806,sarma2014modulation}. The incident optical field ($E$) is the superposition of the forward ($E_f$) and backward ($E_b$) field distributions, and it reads as \cite{erdogan1997fiber}
	\begin{gather}
		E(z)=E_f \exp(ikz)+E_b \exp(-ikz),
		\label{Eq:3}
	\end{gather}
	where $k$ signifies the magnitude of the wave vector. Obtaining the governing equations that describe the propagation of fields in a PTFBG requires the substitution of the squared refractive index given in Eq. (\ref{Eq:2}) and electric field distribution (Eq. (\ref{Eq:3})) in the time-independent Helmholtz equation given below \cite{raja2020phase}:
	\begin{gather}
		\cfrac{d^2E}{dz^2}+k^2\left(\cfrac{n^2(z)}{n_0^2}\right)E=0.
		\label{Eq:4}
	\end{gather} 
	While expanding Eq. (\ref{Eq:4}), derivative terms like $E_f^{''}$ and $E_b^{''}$ can be neglected using the slowly varying envelope approximation (SVEA) \cite{lin2011unidirectional,miri2012bragg}. Along these lines, the rapidly oscillating  exponential terms of the form $\exp[\pm i(2 \pi z/\Lambda+kz)]$ also get neglected \cite{raja2020phase}. The four-wave mixing (FWM) terms $E_f^* E_b$ and $E_b^* E_f$ are assumed to have no significant impact on the propagation \cite{komissarova2019pt,sudhakar2022low,malomed2005coupled}. Under these circumstances, retaining the self-phase modulation (SPM) and cross-phase modulation (XPM) terms is sufficient while expanding the nonlinearity. The ratio between the SPM and XPM terms is 1:1 (mathematically) \cite{malomed2005coupled,merhasin2007solitons}. Also, the equations are further rearranged for the forward and backward propagating fields with retention of the phase mismatch and effective feedback terms. With these assumptions, the resulting equation reads
	
	\begin{gather}
		\nonumber	iE_f^{'}\exp(ikz)-iE_b^{'}\exp(-ikz)\\\nonumber+(\kappa+g)E_b\exp[ i(2 \pi z/\Lambda-kz)]\\\nonumber+(\kappa-g)E_f\exp[-i(2 \pi z/\Lambda-kz)]\\-S\cfrac{E_f \exp(ikz)+E_b \exp(-ikz)}{1+|E_f|^2+|E_b|^2}=0.
		\label{Eq:5}
	\end{gather}
	
	The nonlinear first-order differential equations for the transmitted and reflected waves read as
	\begin{gather}
		\nonumber iE_f^{'}+(\kappa+g)E_b\exp[ i(2 \pi z/\Lambda-2 kz)]\\-\cfrac{S E_f}{(1+|E_f|^2+|E_b|^2)}=0,
		\label{Eq:6}
	\end{gather}
	\begin{gather}
		\nonumber iE_b^{'}+(\kappa-g)E_f\exp[ -i(2 \pi z/\Lambda-2 kz)]\\-\cfrac{S E_b}{(1+|E_f|^2+|E_b|^2)}=0.
		\label{Eq:7}
	\end{gather}
	
	From the fundamentals of FBGs, the detuning parameter ($\delta$) that indicates the deviation in the operating wavelength ($\lambda$) from the Bragg wavelength ($\lambda_b$) reads as $\delta=k-\pi/\Lambda=2\pi n_0\left(\cfrac{1}{\lambda}-\cfrac{1}{\lambda_b}\right)$. Numerically, it is possible to separate the detuning parameter from the coupling term by adopting a transformation $u,v=E_{f,b} \exp(\mp i \delta z)$ \cite{porsezian2005modulational}, and the resulting equations read
	\begin{gather}
		\frac{d u}{dz}=i\delta u+i \left(k + g\right)v-\cfrac{i S u}{(1+ |u|^{2}+|v|^{2})},
		\label{Eq:8}
	\end{gather}
	\begin{gather}
		-	\frac{d v}{dz}=i\delta v+i \left(k - g\right)u-\cfrac{i S v}{(1+ |u|^{2}+|v|^{2})}.
		\label{Eq:9}
	\end{gather}
	
	These equations are valid for left light incident conditions. Under a reversal in the direction of light incidence (right), the term $\kappa + g$ in Eq. (\ref{Eq:8}) changes to $\kappa - g$. Similarly, the term $\kappa - g$ in Eq. (\ref{Eq:9}) is replaced by $\kappa + g$.
	
Before delving into the theoretical investigation of the system based on the governing model, it is crucial to address why its consideration holds merit. To do so, we must first highlight how this mathematical model distinguishes itself from those already discussed in the existing literature \cite{malomed2005coupled,merhasin2007solitons}. Previous studies in the literature focus on exploring the dynamics of solitons and their interesting collision properties in photorefractive crystal-based one-dimensional optical lattices and bulk longitudinal photo-induced gratings. Although the fundamental study pertaining to the stable nature of solitons with phase-matched conditions has been quite extensively investigated in these systems \cite{malomed2005coupled,merhasin2007solitons}, both the fundamental and the application perspectives of continuous wave (CW) remain unexplored to date. The steady-state model presented in this article, using ordinary differential equations (ODEs) in the presence of CW input, is instrumental in studying various aspects of bistability and multistability. This includes the shape of hysteresis curves, hysteresis width, and their dependence on grating parameters such as gain/loss, and direction of light incidence. The same system can also be investigated using partial differential equations (PDEs) with pulsed input. Although the switching thresholds generated by each model may differ, the overall switching behavior of each hysteresis curve may qualitatively remain the same. However pulses may exhibit more stable nature than that of the CW inputs in such nonlinear systems because of their intrinsic characteristics. Nonetheless, detailed numerical investigations are required to fully understand the system's behavior, which we hope to address separately in future. In the present investigation, we analyze the switching properties of CW states in the FBGs with an additional term known as the detuning parameter indicating a substantial difference between the Bragg wavelength and the wavelength of the input light.

	The inclusion of $\mathcal{PT}$-symmetry terms sets the current governing model apart from those found in the literature, presenting a distinctive feature that diverges from established formulations and introducing a unique dimension to the theoretical framework, enabling the study of OB in diverse operating conditions, including both unbroken and broken regimes, under two different light incidence conditions (left and right). This multifaceted approach, made possible by the presence of $\mathcal{PT}$ symmetry terms, is notably impossible in existing models where the absence of such terms limits the feasibility of exploring such scenarios. Similarly, the incorporation of the detuning parameter into the system further distinguishes the present model from existing ones, enabling the study of nonlinear characteristics at non synchronous wavelengths. This addition not only enhances the practicality of the approach at phase mismatched conditions but also provides an additional degree of freedom to manipulate the characteristics of OB/OM curves.

	We use the well-known implicit Runge-Kutta fourth-order method to solve the system of coupled equations in (\ref{Eq:8}) and (\ref{Eq:9}) with the following boundary conditions
	\begin{gather}
		u(0)=u_0 \quad \text{and} \quad v(L)=0.
		\label{Eq:10}
	\end{gather}
	The input and output intensities read as $P_0 = |u_0|^2$ and $P_1(L)=|u(L)|^2$, respectively.
	
	The nonlinear PTFBG works in the unbroken $\mathcal{PT}$-symmetric regime under the condition $\kappa > g$ \cite{miri2012bragg,sarma2014modulation}. At the unitary transmission point, an inevitable phenomenon is the possibility of the breaking of the $\mathcal{PT}$-symmetry in the system, when $\kappa = g$. Above this condition, the system operates in the broken regime where \textcolor{blue}{$\kappa < g$}.  An alternative perspective on the boundary of $\mathcal{PT}$-symmetry breaking can be gained by investigating the dispersion curves supported by the same system (but with partial differential equations taking into account the time co-ordinate), which was recently done by Tamilselvan \emph{et al.} \cite{tamilselvan2023modulational} in the case of modulational instability analysis.   We investigate the proposed system for different values of NL parameters at two different lengths ($L = 20$ and 70 $cm$). 
	\subsection{Organization of the results}
	 Section \ref{Sec:3} -- deals with the transmission characteristics of conventional FBG with SNL. The gain-loss parameter serves as an additional degree of freedom to manipulate the OB curves. Further, it allows us to study the switching in two different $\mathcal{PT}$-symmetric operating regimes, namely, the unbroken and broken $\mathcal{PT}$-symmetric regimes. Sections \ref{Sec:4} and \ref{Sec:5} deal with the input-output characteristics of the proposed system in the unbroken and broken regimes, respectively.    
	
In the present work, the system parameters consistently work in conjunction with the SNL parameter, resulting in random variations in the OB and OM curves even for the slight changes in their values. The peculiar OB and OM curves in the present work differ significantly from existing systems studied in the literature. This underscores the importance of conducting a comprehensive study of both regimes to understand their responses to variations in system parameters. Hence, Secs. \ref{Sec:4} and  \ref{Sec:5} are divided into multiple subsections, based on the nature of the hysteresis curves.	The system manifests three types of OM curves: S-shaped, ramp-like, and mixed. S-shaped OB curves demonstrate gradual changes in output with input intensity tuning, while ramp-like OB and OM curves exhibit sharp variations. Mixed OM curves blend characteristics of ramp-like and S-shaped OM, showing sharp changes at low intensities and gradual changes at high intensities. The hysteresis width decreases with input intensity tuning for S-shaped OM and increases for ramp-like OM. Each curve is explored separately to understand its properties and parameter-induced changes. Note that the novel and different OB and OM curves described in the manuscript are unique and not previously documented
in the literature. Classifying and naming these different types of curves provide a standardized
framework for describing and discussing them within the scientific community as each shape corresponds to specific family of solutions of the ODEs.  Also, different applications including the construction of logic gates may require specific types of OB and OM curves for an optimal performance. For instance, the ramp-like OB/OM may aid in ultra-fast switching owing to their sharp response to the input power as opposed to the standard S-shaped OB/OM curves.

Section  \ref{Sec:6} recalls the important results of the present work. 
	\subsection{Overview of OB}
	Before we present the simulation results, we provide an overview of the OB and OM phenomena. As the input intensity ($P_0$) varies, the output intensity [$P_1(L)$] increases linearly. A sudden jump in the output intensity occurs at one particular value known as switch-up intensity ($P_{th}^{\uparrow}$), indicating an onset of a second stable branch of the OB, and the mechanism is commonly known as switch-up action. A part of the input-output curve corresponding to input intensities lying between zero and switch-up represents ($0<P_0<P_{th}^{\uparrow}$)  the first stable branch of the OB curve. Once the system switches to the second branch, output remains in it for a given range of input intensities. In the case of OM, switching to the successive stable states happens at distinct switch-up intensity values.  Tuning the input intensity in the reverse direction completes the s-shaped hysteresis curve. During this process, the system does not return to the previous stable branch at the same switch-up intensity value.  However, it returns to its previous state at another critical intensity known as the switch-down intensity ($P_{th}^{\downarrow}$) during the switch-down mechanism. For any value of input intensities between the switch-up and down values ($P_{th}^{\downarrow}<P_0<P_{th}^{\uparrow}$), the output of the system is bistable. The difference between the critical switch-up and down intensities dictates the width of the hysteresis curve ($\Delta P_{th}$ = $P_{th}^{\uparrow} - P_{th}^{\downarrow} $).

		\subsection{Some practical considerations}
	It is important to emphasize that although observing the OB and OM curves in FBGs is possible experimentally, it requires addressing several practical challenges, including the identification of a suitable fiber material that offers SNL at relatively low power.  From the available scientific literature, it is evident that semiconductor-doped glass having strong SNL can possess a third-order nonlinear coefficient ($n_2$) in a range varying from $10^{-15}$ to $10^{-13}$ $m^2/W$ \cite{ironside1988nonlinear,hall1989nonlinear}.  $CdS_{1-x}Se_{x}$ is an example of this type of glass material characterized by a fast nonlinear response time of $10^{-11}$ seconds and a significant nonlinearity value \cite{ironside1988nonlinear,jain1983degenerate,acioli1988measurement,roussignol1987new}.  With the value of $A_{eff}$ assumed to be 100 $\mu m^2$, the cubic nonlinearity ($S$) is found to be 0.5927 $cm/W$ at the operating wavelength of 1060 $nm$ and $n_2 = 10^{-15} m^2/W$ \cite{coutaz1991saturation,ironside1988nonlinear,kang1995femtosecond} (note that a conventional silica fiber, on the other hand, exhibits a third-order nonlinearity value of $2.6 \times 10^{-20}$ $m^2/W$ \cite{agrawal2000nonlinear,yosia2007double,ping2005bistability,ping2005nonlinear})
		
	Along these lines, the third-order susceptibility [$\psi^{(3)}$] of sulfide and heavy-metal doped oxide glass with refractive index 2.19 -- 2.5 lies, respectively, within a range of 3.1 -- 5.6 $\times$ $10^{-13}$ esu at 1.06 $\mu$m \cite{kang1995femtosecond,borrelli1995resonant} and 1.2$\pm$0.4 -- 7.9$\pm$2.4 $\times$ $10^{-13}$ esu at 1.25 $\mu$m \cite{kang1995femtosecond}. For instance, the third-order nonlinear coefficient [$n_2 = 24 \pi/n_0$ $\psi^{(3)}$] of $Pb_O$(60)$TeO_2$(25)$SiO_2$(15) at 1.25 $\mu$m measures to be 3.322 $\times$ $10^{-12}$ $m^2/W$, provided that the refractive index is 2.27 and $\psi^{(3)} = 1 \times 10^{-13}$ \cite{kang1995femtosecond,borrelli1995resonant}.
		
		  Some other examples of this type of glass materials are $GeS_{2}(87.3)Ga_{2}S_{3}(13.7)$, $PbO(60)TeO_{2}(25)SiO_{2}(15)$, and $La_2S_3(35)Ga_2S_3(65)$ \cite{kang1995femtosecond}.  In light of these facts, we present some practical values that can aid experimental physicists in creating low-power OB/OM curves influenced by SNL. These values are listed in Table \ref{tab7}. For comparative purposes, the values that were used for the formation of OB and OM curves are also presented from the already existing literature.
	
	\begin{table*}
		\caption{\centering{Comparison of various device parameters used in physical units }}
		\begin{center}
			\begin{tabular}{c c c c }
				\hline
				\hline
				{Symbol}&{Device } & {physical} & {values} \\
				{}&{parameter} & {values} & {used in the literature} \\
				\hline
				
				{$n_2$}&{third-order nonlinear}&{$10^{-15}$ -- $10^{-13}$ $m^2/W$ \cite{ironside1988nonlinear,kang1995femtosecond}}&{$2.6 \times 10^{-20}$ $m^2/W$ \cite{lee2003nonlinear,ping2005bistability,shum2007optical}}\\
				{}&{coefficient}&{}&{2.2 $\times 10^{-16}$ $m^2/W$ \cite{ping2005nonlinear}}\\
				{}&{}&{}&{2.7 $\times 10^{-13}$ $cm^2/W$ \cite{yosia2007double}}\\
				{}&{}&{}&{2 $\times 10^{-17}$ $m^2/W$ \cite{phang2013ultrafast}}\\
				{}&{}&{}&{}\\
				
				{$A_{eff}$}&{effective area of}&{100 $\mu m^2$ }&{1 -- 100 $\mu m^2$ \cite{agrawal2000nonlinear}}\\
				{}&{the fiber}&{}&{}\\
				{}&{}&{}&{}\\
				{}&{}&{}&{}\\

				{$S$}&{saturable nonlinearity}&{0.5 -- 6 $cm/W$}&{1.0544 $\times$ $10^{-5}$ $m/W$ \cite{lee2003nonlinear,ping2005bistability,shum2007optical}}\\
				{}&{}&{}&{}\\
				{}&{}&{}&{}\\

				{$L$}&{device length}& {20 $cm$} &{1 $cm$ \cite{ping2005bistability,ping2005nonlinear,yosia2007double,shum2007optical}, 3.5 $cm$ \cite{lee2003nonlinear}}\\
				{}&{}&{}&{}\\

				{$\kappa$}&{coupling coefficient}& {0.4 $cm^{-1}$} &{0.8 $cm^{-1}$ \cite{lee2003nonlinear}, 5 $cm^{-1}$\cite{ping2005bistability,shum2007optical}}\\
				{}&{}&{}&{ }\\
				{}&{}&{}&{} \\

				{$g$}&{gain-loss }& {0 (conventional)} &{}\\
				{}&{coefficient}&{0 -- 4 $cm$ $^{-1}$}&{0 - 1200 $cm$ $^{-1}$ \cite{phang2015versatile, phang2014impact}}\\
				{}&{}&{(unbroken regime)}&{} \\
				{}&{}&{}&{} \\
				
				{$n_0$}&{refractive index of the core }& {2.27} &{2.19 -- 2.5 \cite{kang1995femtosecond,borrelli1995resonant}}\\
				{}&{}&{}&{}\\
				
				{$\lambda_b$}&{Bragg wavelength} & {1060 nm} &{1000 nm \cite{phang2013ultrafast,phang2014impact}} \\
				{}&{}& {} & {}\\
				
				{$\lambda_b$$-$$\lambda$}&{variations in operating} & {$\pm$ 0.05 nm}&{-0.015 nm \cite{yosia2007double}, 0.125 nm \cite{shum2007optical,ping2005bistability}} \\
				{}&{ wavelength ($\lambda$) from $\lambda_b$}& {} &{}\\
				{}&{}& {}  &{}\\

				{$\delta$}&{detuning parameter}& {0 -- 2.5 $cm^{-1}$} &{0.005 $cm^{-1}$ \cite{lee2003nonlinear}, -0.9611 $cm^{-1}$ \cite{ping2005bistability},  }\\
				{}&{}&{}&{-1.8039 $cm^{-1}$ \cite{ping2005nonlinear}, 4.74 $cm^{-1}$ \cite{shum2007optical},  }\\
				{}&{}&{}&{4.75 $cm^{-1}$ \cite{lee2003nonlinear}}\\
				{}&{}&{}&{}\\

				{$P_0$ and $P_1(L)$ }&{input and output }&{0 -- 15 $MW/cm^2$,}&{0 -- 400 $MW/cm^2$ \cite{yosia2007double},}\\
				{}&{intensities}&{}&{0 -- 100 $GW/cm^2$ \cite{lee2003nonlinear}}\\
				{}&{}&{}&{0 -- 30 $GW/cm^2$ \cite{shum2007optical,ping2005bistability}}\\
				{}&{}&{}&{0 -- 4 $\times$ $10^{14}$ $W/m^2$ \cite{phang2014impact,phang2015versatile}}\\
				
				\hline\hline
			\end{tabular}
			\label{tab7}
		\end{center}
	\end{table*}

We chose the Bragg wavelength in the numerical studies to be 1060 nm because the existing literature predominantly covers experimental studies on the nonlinear-optical effects of heavy-metal and sulfide-glass at near-infrared wavelengths between 1 and 1.250 $\mu$$m$ \cite{borrelli1995resonant,kang1995femtosecond}. Given an assumed refractive index of 2.27 for the background material [$Pb_O(60)TeO_2(25)SiO_2(15)$], a difference of $0.05$ $nm$ between the operating wavelength of the system and $\lambda_b$ can induce a detuning value ($\delta$) of 6.3472 $cm^{-1}$ \cite{kang1995femtosecond}.
	
Similarly, another essential experimental challenge is finding a suitable CW source that can deliver a high $kW$ power in the given wavelength range. YLR CW Ytterbium fiber lasers that are available (commercially) are highly suitable for this purpose because of their high stability, efficiency, beam quality, and long lifetime. They can provide an input power of up to 4 $kW$ to the system \cite{CWsource}. It is important to note that intensity is a measure of power confined per unit area of the core. For instance, if the effective area of the core is 100 $\mu$$m^2$ and the input power is in Watts, the input intensities ($P_0$) that are required to create OB and OM states can be in the order of $MW/cm^2$ \cite{shum2007optical, phang2014impact, phang2015versatile}. Developing PTFBGs with complex refractive index profiles can be accomplished through external pumping in fibers doped with rare-earth dopants. Nonetheless, selecting a suitable dopant material that can generate the desired gain and loss regions is a notable challenge. To overcome this obstacle, adding $Er^{3+}$ and $Cr^{3+}$ dopants onto a suitable glass substrate to develop gain and loss regions, respectively, is a promising option. 

\begin{figure*}[t]
	\centering	\includegraphics[width=0.35\linewidth]{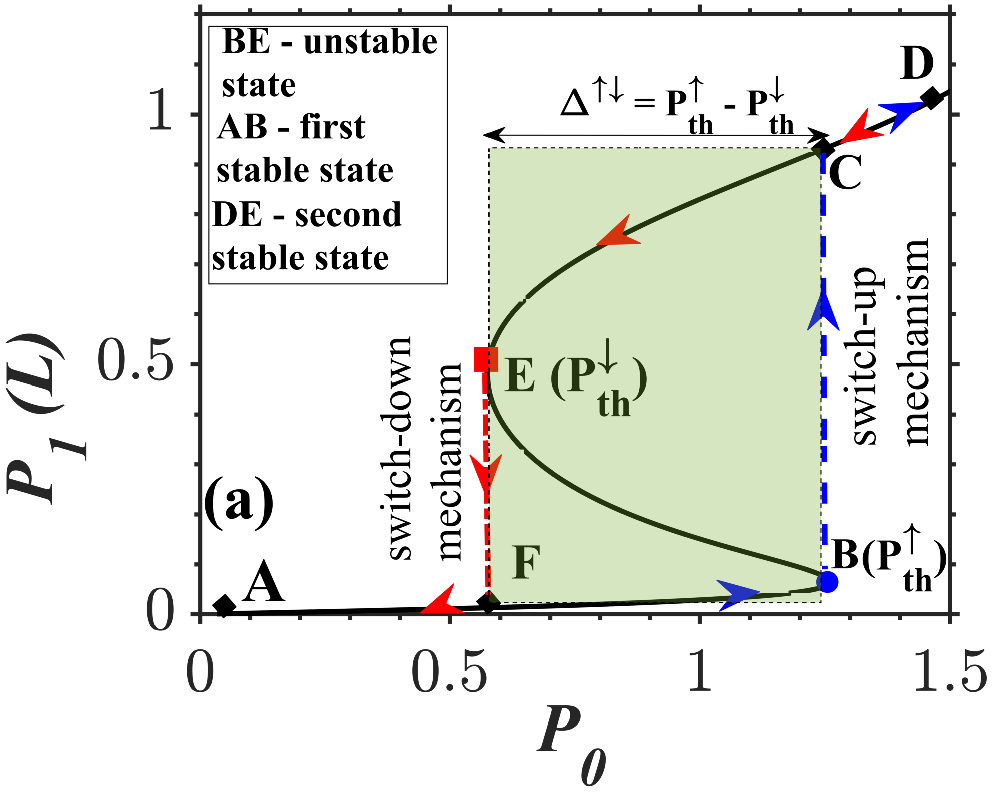}\includegraphics[width=0.35\linewidth]{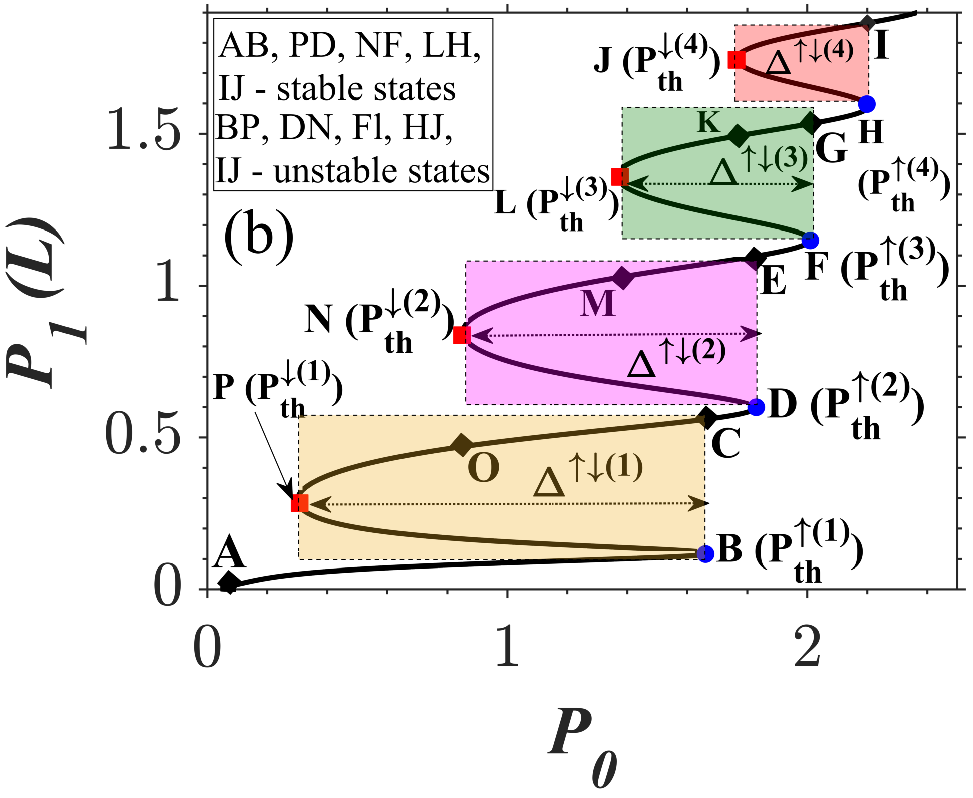}
	\\\includegraphics[width=0.35\linewidth]{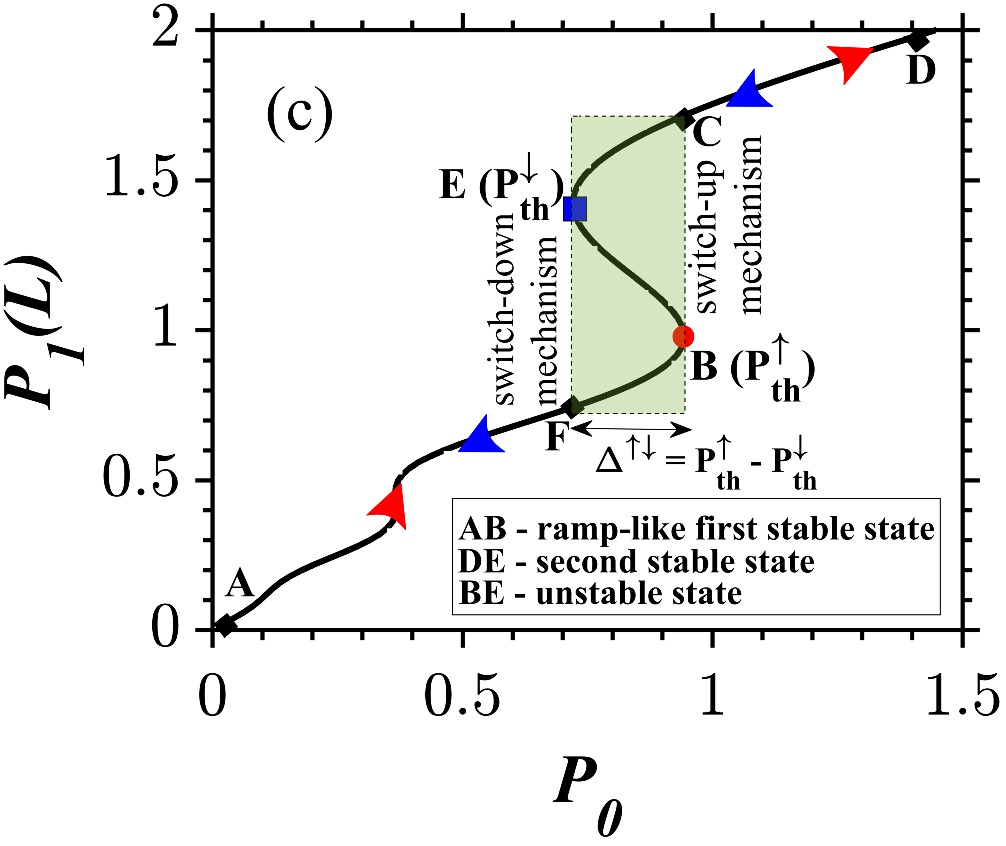}\includegraphics[width=0.35\linewidth]{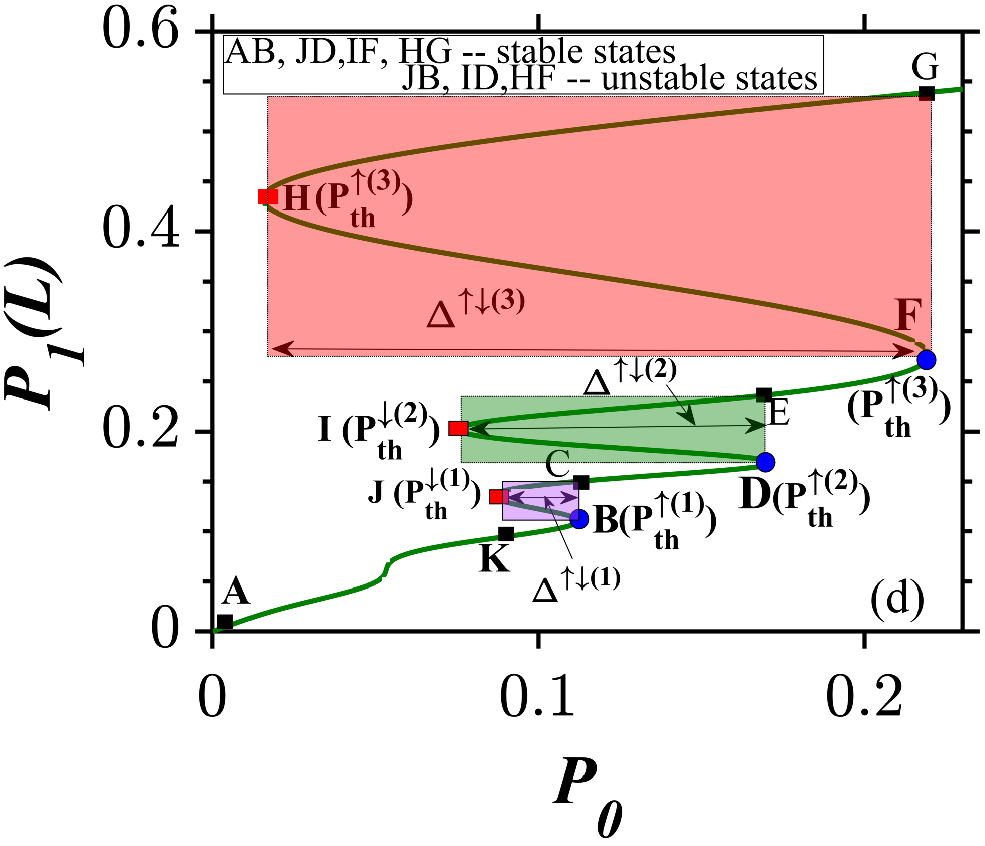}\\\includegraphics[width=0.35\linewidth]{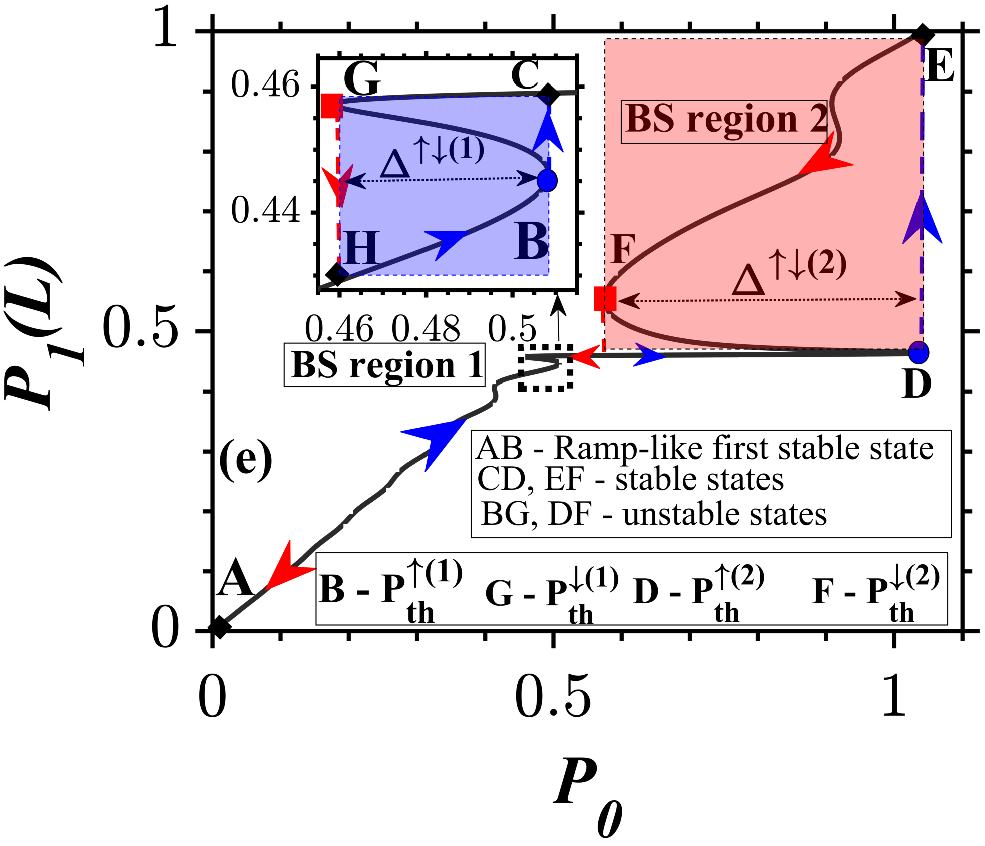}
	\caption{Schematic sketches showing different stable and unstable
		regions of typical (a) S-shaped OB, (b) S-shaped OM, (c) ramp-like OB, (d) ramp-like OM, and (e) mixed OM curves in a PTFBG with SNL. The S-shaped OB and OM curves feature gradual variations in the output intensities against the input intensities, as shown
		in (a) and (b). In the case of S-shaped OM, the hysteresis width of successive stable branches decreases with an increase in the input intensity, as shown in (b). Plot in (c) depicts the
		ramp-like OB featuring a sharp variation (as opposed to the gradual variation in the S- shaped OB curve, in particular in the first stable state where the S- shaped OB exhibits a flat variation) in the output intensity against the increase in the input intensities. In the case of ramp-like OM, the width of the successive stable branches decreases, as observed in (d). Mixed OM curve is a fusion between the ramp-like and
		S-shaped OM curves, as delineated in (e). At low intensities,
		the variations in the output against input are sharp
		(a typical feature of ramp-like OM). However, the output intensity
		varies gradually against the input intensity at higher intensities
		(a ubiquitous feature of the S-shaped OM curve). In
		the case of mixed OM curves, the plots feature two distinct regions,
		where Regions I and II represent the variations in the
		OM curves at low and high intensities, respectively. Note that
		in Region I (Region II), the hysteresis width of the successive
		stable branch increases (decreases) with an increase in the
		input intensity.}
	\label{fig0}
\end{figure*}

\begin{figure}[t]
	\centering	\includegraphics[width=0.45\linewidth]{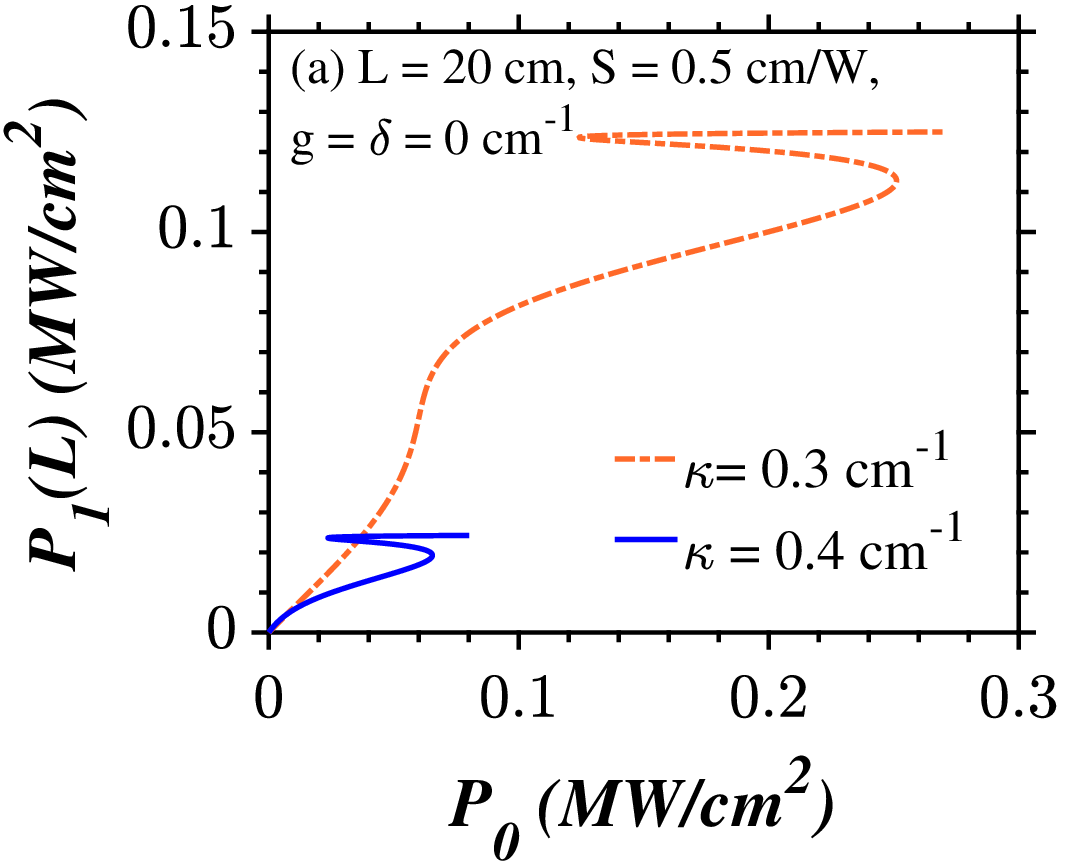}\includegraphics[width=0.45\linewidth]{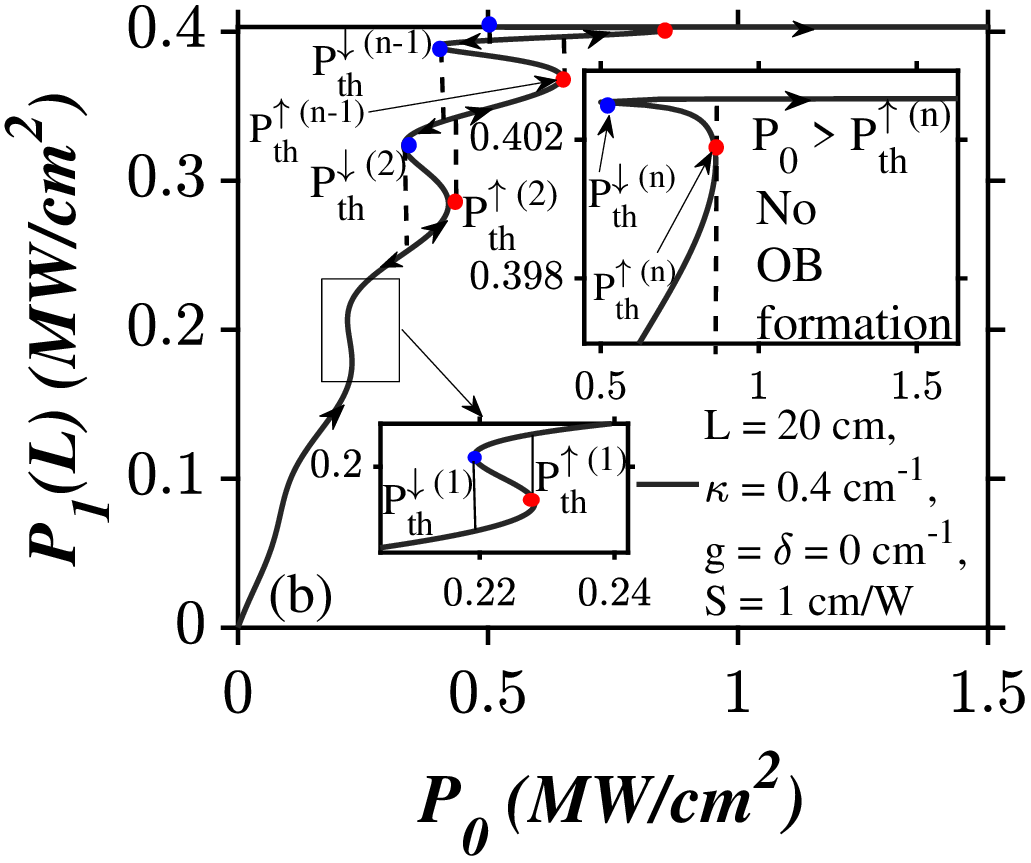}
	\\\includegraphics[width=0.45\linewidth]{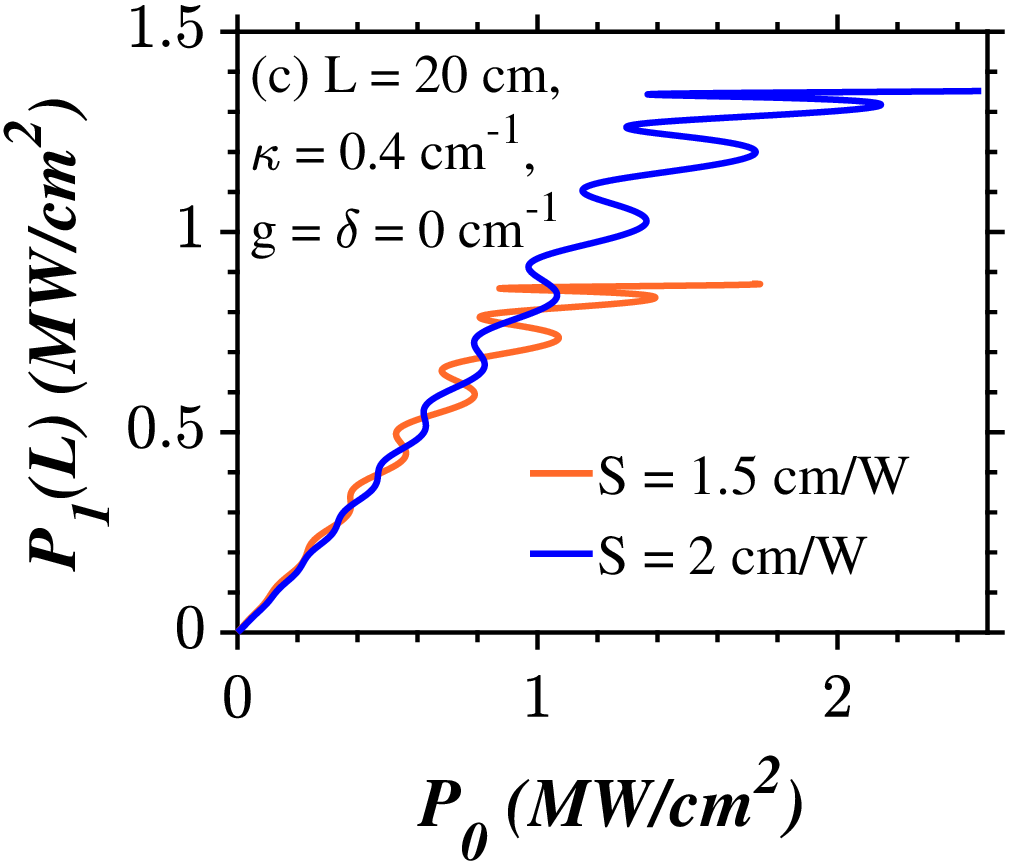}\includegraphics[width=0.45\linewidth]{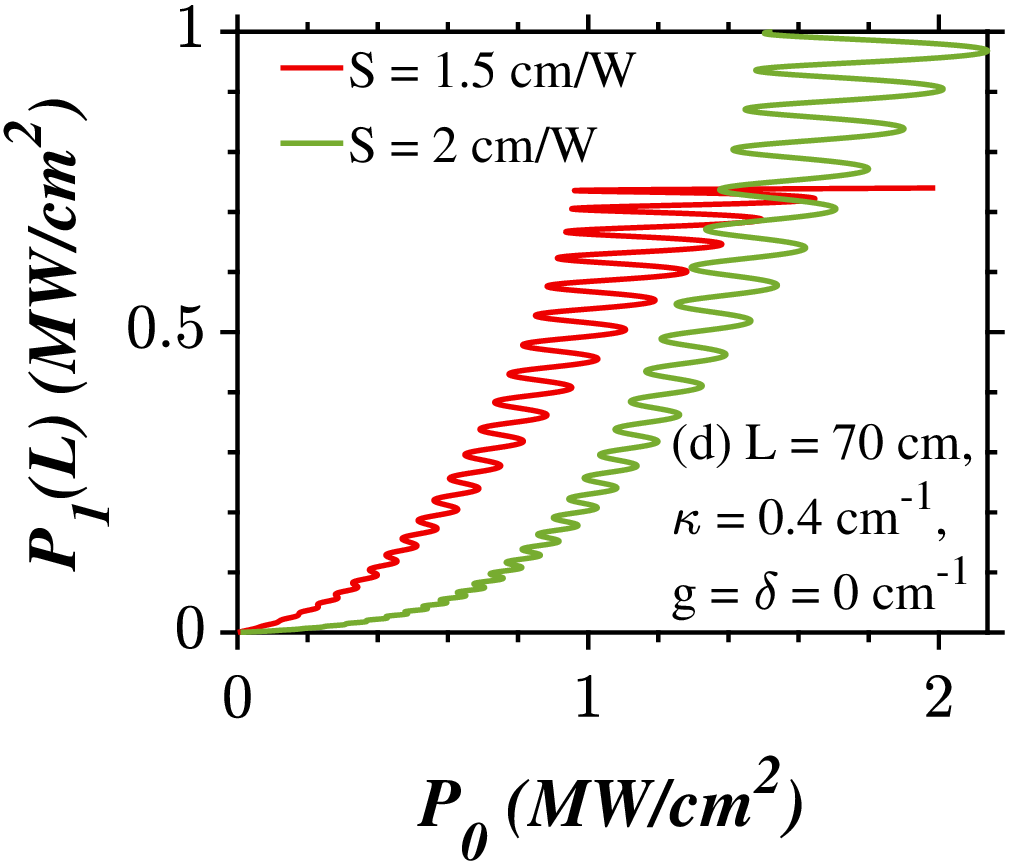}
	\caption{OB (OM) curves exhibited by a conventional FBG with SNL at $\delta$ = 0 $cm^{-1}$. Role of (a) coupling coefficient ($\kappa$), (b), (c) SNL ($S$), and (d) device length ($L$) on the OB (OM) curves. Note that here and in upcoming figures, the values of system parameters are provided within the plots.}
	\label{fig1}
\end{figure}

	\section{OB/OM in conventional Bragg structures with SNL}
	\label{Sec:3}

	We first investigate the role of the coupling parameter on the input-out characteristics of the bistable states. As the input intensity varies from zero upwards, the output intensities increase sharply, leading to a ramp-like first stable state, as shown in Fig. \ref{fig1}(a).  The output jumps from the first to the second stable state at the switch-up intensity and remains in it for a finite increase in input intensities, as delineated in Fig. \ref{fig1}(a). 	It is well-known that the feedback offered by the system is an essential ingredient besides the intensity-dependent refractive index for the OB/OM to occur in FBGs. Insufficient coupling strength inhibits the formation of OB, say $\kappa < 0.3$ $cm^{-1}$. The switch-up and down intensities decrease with an increase in the coupling parameter ($\kappa$), as shown in Figs. \ref{fig1}(a).  Additionally, the hysteresis width increases with an increase in the coupling coefficient. On the other hand, the value of $\kappa$ cannot be arbitrarily large in PTFBGs \cite{raja2019multifaceted,PhysRevA.100.053806}.    Optimizing the device length ($L$) is essential for the desirable OB/OM curves to appear.  In practice, the value of the coupling parameter ($\kappa$) ranges from 1 to 10 $cm^{-1}$ \cite{agrawal2001applications}. In the literature, we could find FBGs fabricated in a wide range of physical lengths ranging from 1 $mm$ to 20 $cm$ \cite{broderick2000nonlinear}. In our numerical experiments, we observe that desirable OB curves in the input-output characteristics of a FBG with SNL occur when the coupling coefficient reduces to a value approximately ten times less than the values we used in our previous works \cite{raja2019multifaceted, PhysRevA.100.053806}. Therefore,  the coupling coefficient ($\kappa$) is assumed to have a value of 0.4 $cm^{-1}$ throughout this article (unless specified).  However, the product of these two parameters ($\kappa L$) will never go beyond the permitted numerical values (1 to 100) as the reduction in the coupling gets compensated by the increment in the device length \cite{raja2019multifaceted}. With this note, we now look into the nonlinear transmission characteristics of the system under different operating conditions. 
	
	As we tune the nonlinearity parameter gradually, a transition from the ramp-like OB ($S < 1$ $cm/W$) to the ramp-like OM  ($S \ge 1$ $cm/W$) is visible in Fig. \ref{fig1} (b). The number of stable states in the ramp-like OM curve increases with an increase in the NL, as shown in Fig. \ref{fig1}(c). An increase in the NL parameter ($S$) increases the switch-up and down intensities of the OM curve.  In the Figs. \ref{fig1}(a) -- (c), we find that the number of stable states is less for smaller device lengths.  We can generate ramp-like OM curves with more stable states by tuning the device length in the simulations, as shown in  Fig. \ref{fig1}(d). In other words, the higher the value of $L$, the higher the number of stable states.   
	
	At this juncture, we wish to emphasize that generally, nonlinear FBGs display an S-shaped hysteresis curve in their input-output characteristics \cite{winful1979theory,karimi2012all,lee2003nonlinear,ping2005bistability,radic1994optical,radic1995analysis,radic1995theory}. On the contrary, the input-output characteristics of the proposed system display ramp-like OB and OM curves, which were observed only in the broken $\mathcal{PT}$- symmetric systems in our earlier studies \cite{raja2019multifaceted}. Nevertheless, in the literature, these kinds of ramp-like OB and OM curves have been observed in other physics settings (other than different FBGs) such as coupled active ring resonators \cite{zhang2012novel} and photonic metamaterial multilayers with graphene sheets \cite{wen2020tunable}, graphene surface plasmons \cite{da1995dynamics}, plasmonic multilayer nanoparticles \cite{daneshfar2017switching}, plexcitonic systems \cite{naseri2018optical}, graphene-coated nanoparticles \cite{naseri2018terahertz}.  The remarkable aspect of the proposed system is the simultaneous existence
		of ramp-like, S-shaped, and mixed OM curves  within the same operating regime, distinguishing it from other nonlinear FBG systems discussed in the literature.
	
	\section{OB in the unbroken $\mathcal{PT}$-symmetric regime}
	\label{Sec:4}
	
	\subsection{Ramp-like OB/OM curves}
	\label{Sec:4A}
	\begin{figure}[t]
		\centering	\includegraphics[width=0.5\linewidth]{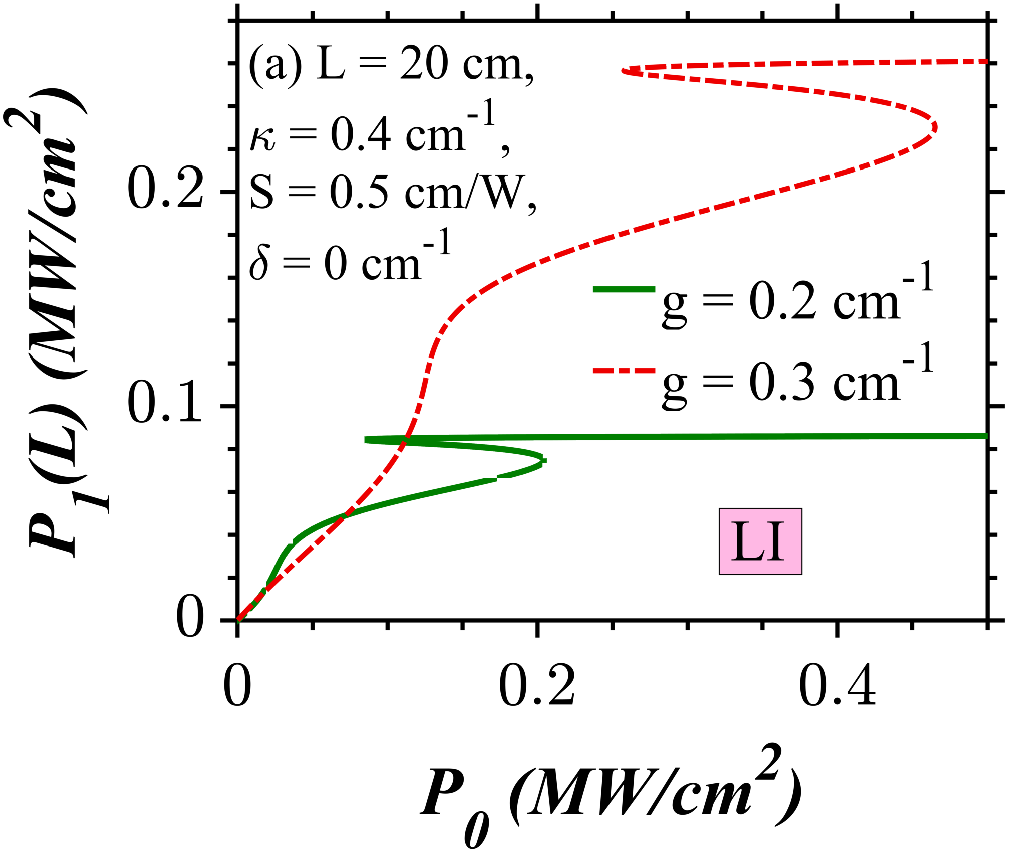}\includegraphics[width=0.5\linewidth]{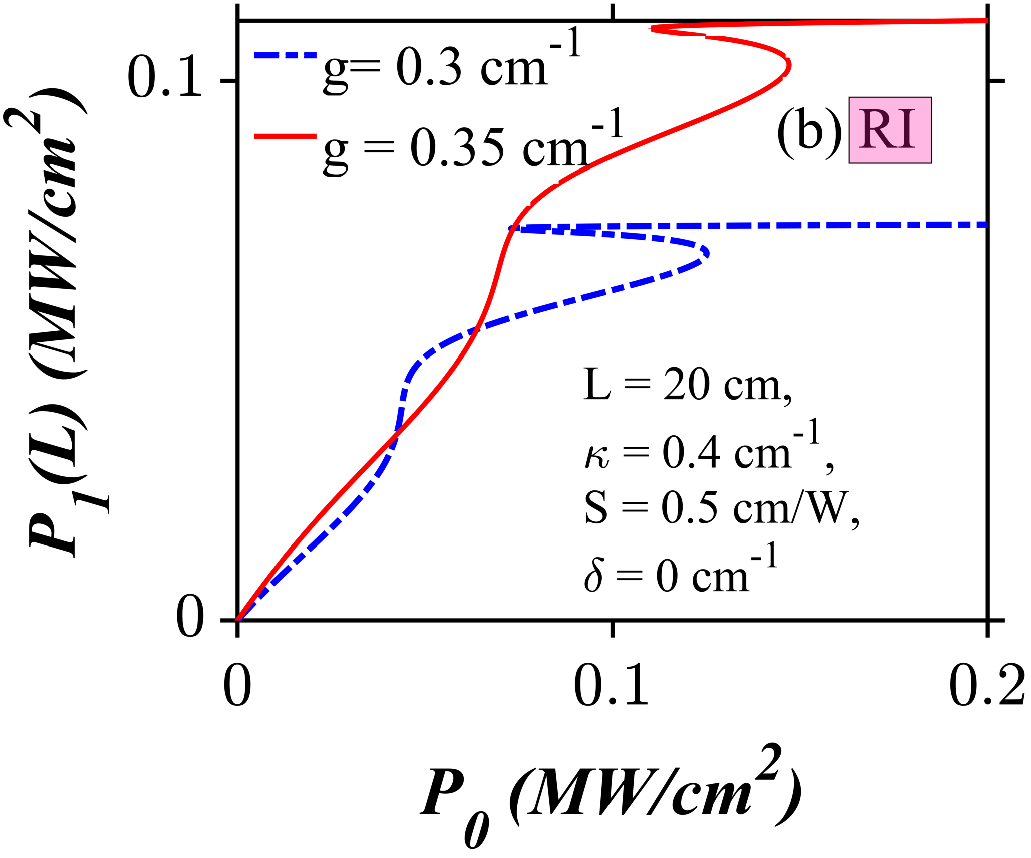}\\	\includegraphics[width=0.5\linewidth]{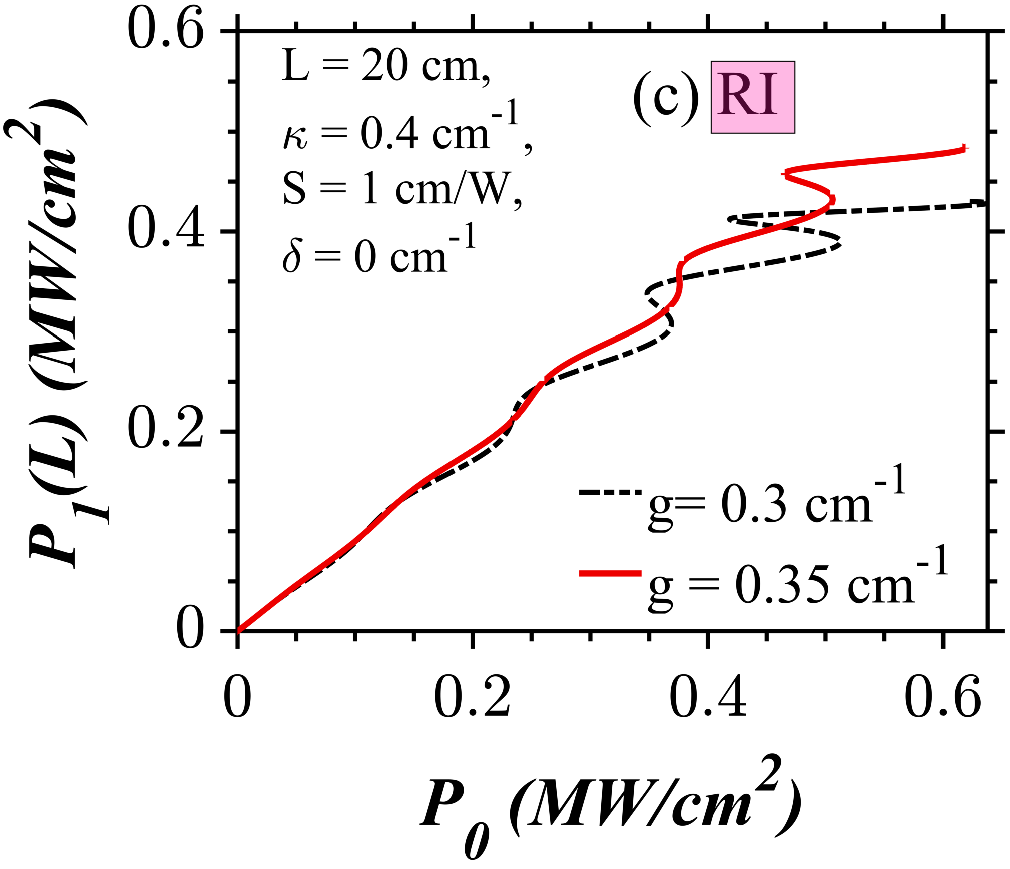}\includegraphics[width=0.5\linewidth]{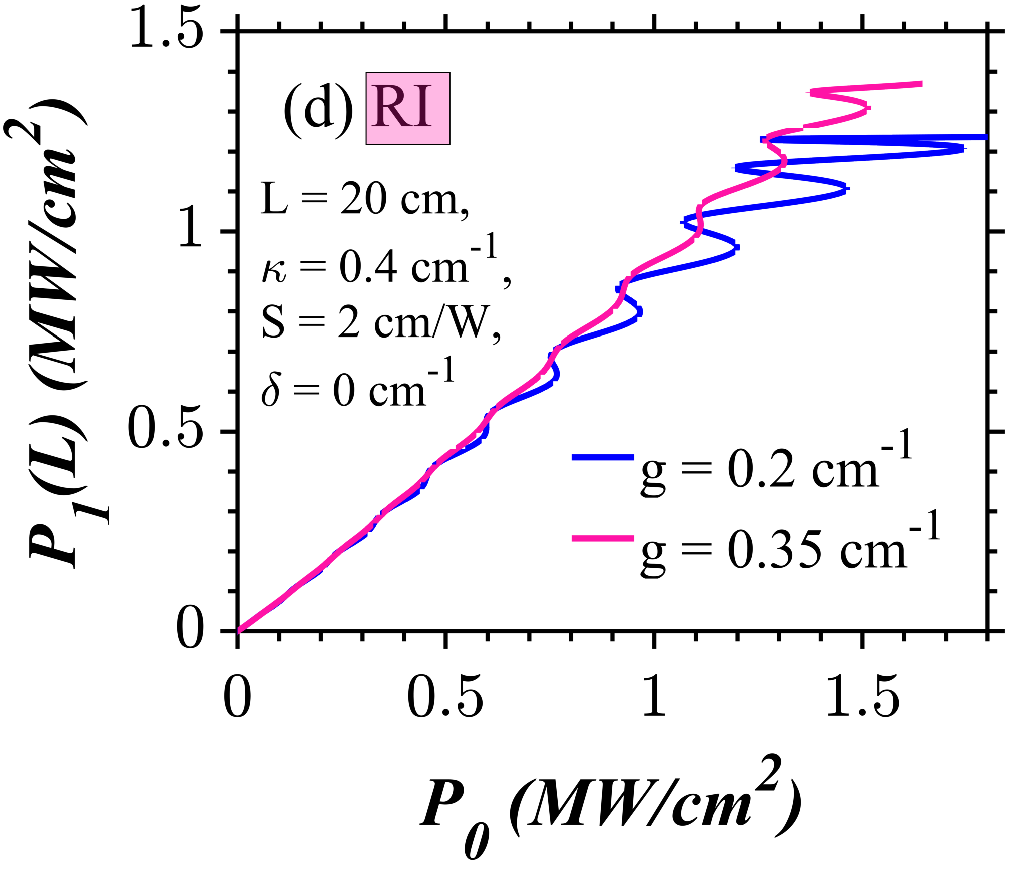}
		\caption{(a) -- (d) Variations in the ramp-like OB (OM) curves shown by PTFBGs with SNL in the unbroken $\mathcal{PT}$- symmetric regime against the gain and loss parameter ($g$) at $\kappa = 0.4$ $cm^{-1}$ and $L = 20$ $cm$ for different values of the SNL parameter ($S$). The light incidence direction is left in (a) and is right in (b) -- (d). }
		\label{fig2}
	\end{figure}
	\begin{figure}[b]
		\centering	\includegraphics[width=0.43\linewidth]{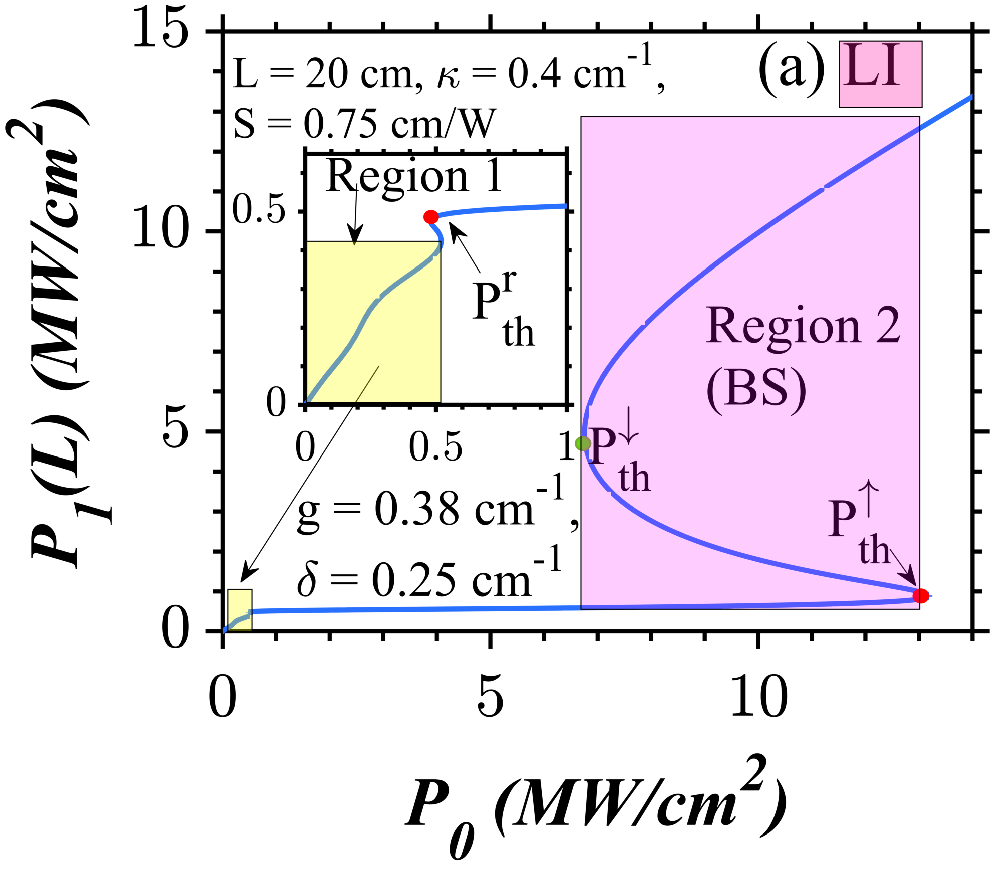}\includegraphics[width=0.43\linewidth]{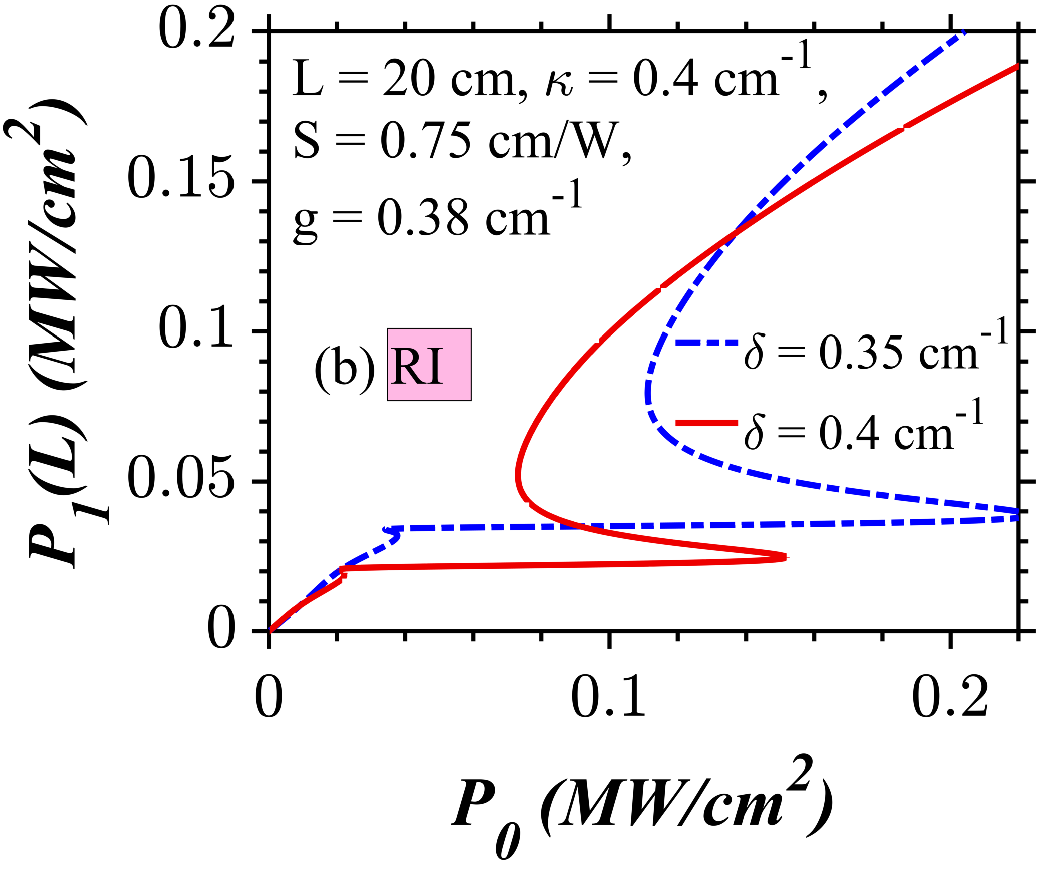}
		\\\includegraphics[width=0.43\linewidth]{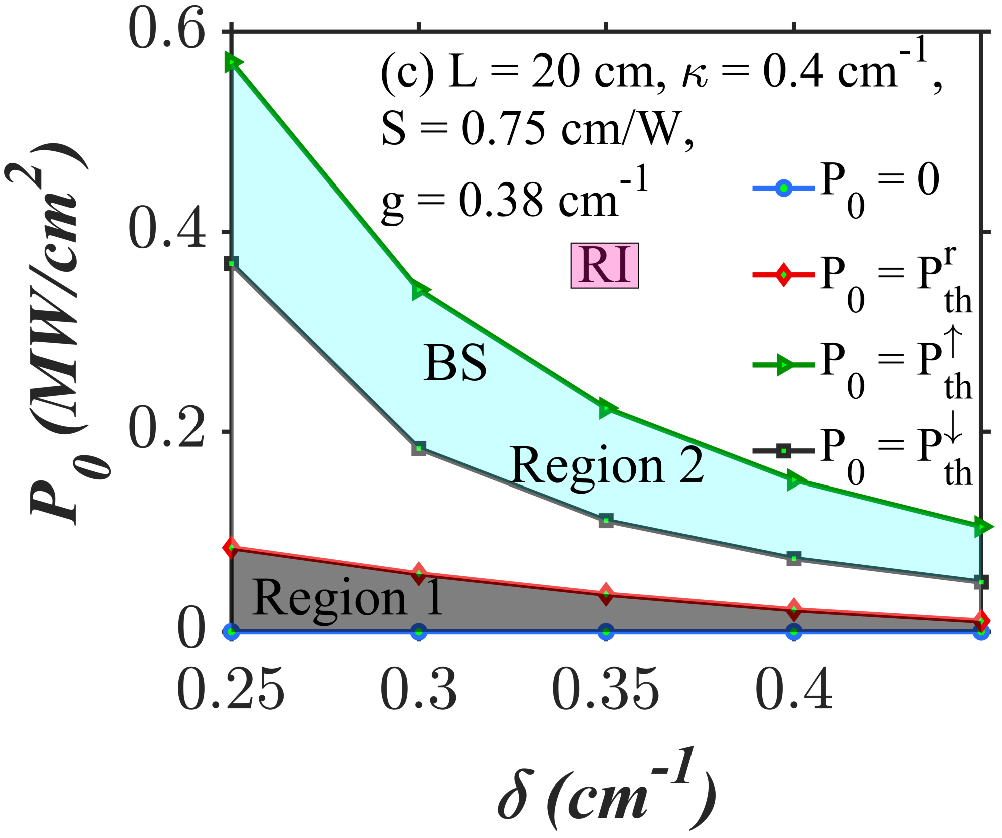}\includegraphics[width=0.43\linewidth]{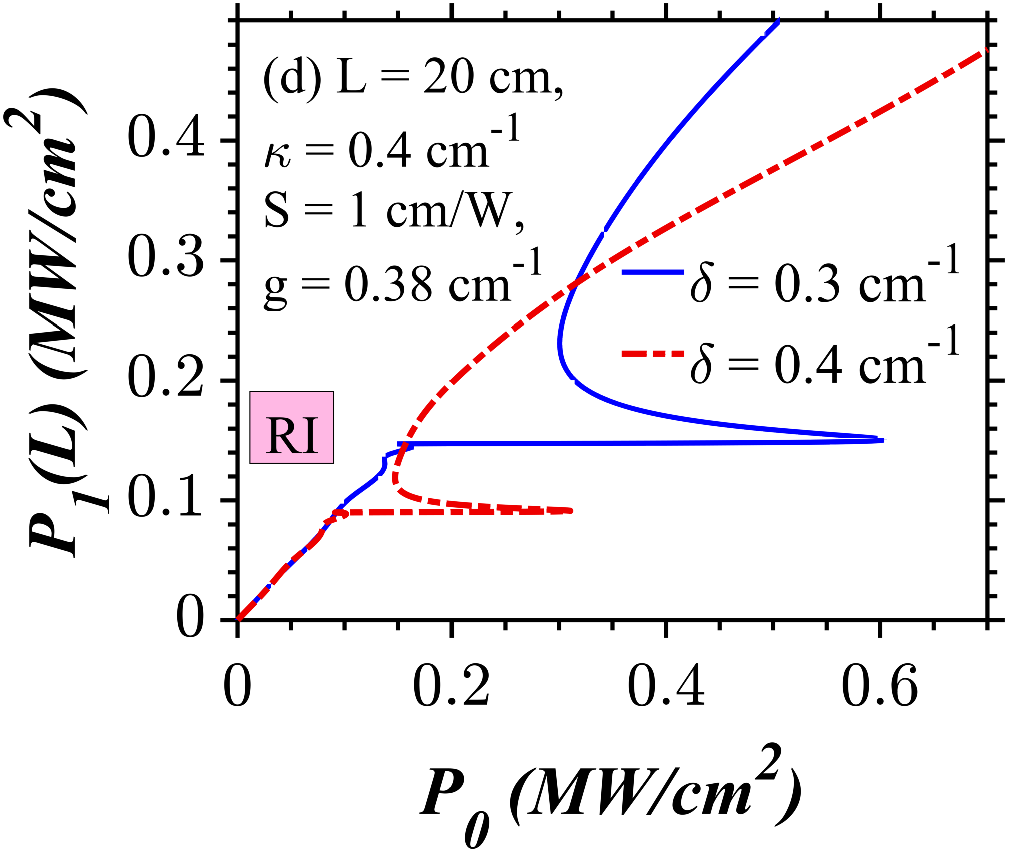}
		\caption{Mixed OB (OM) curves in an unbroken PTFBG with SNL at $\kappa = 0.4$ $cm^{-1}$ and $L = 20$ $cm$. The light launching direction is left in (a) and is right in (b) -- (d).  (c) Variations in the range of input intensities at which ramp-like stable-state appears, and the bistable region (BS) region in (b) against frequency detuning.}
		\label{fig3}
	\end{figure}
 In this section, we present the OB (OM) behavior induced by the impact of $\mathcal{PT}$-symmetry in the unbroken regime. As we tune the input intensities, the system's output varies sharply along the ramp-like first stable branch of a ramp-like OM curve, as shown in Fig. \ref{fig2}(a). In Fig. \ref{fig2}(a), the switching intensities of the ramp-like OM curve are high, which decreases under the reversal in the direction of light incidence, as shown in Fig. \ref{fig2}(b). As a consequence of the increase in the NL parameter, the ramp-like OB ($S < 1$ $cm/W$)  curves transform into ramp-like OM ($S \ge 1$ $cm/W$), as shown in Figs. \ref{fig2}(c) and (d). The width of each hysteresis curve is broader than its former in these ramp-like OM  curves.

	\subsection{Mixed OM curves}
	\label{Sec:4C}

	  In the previous sections, we observed that FBGs and PTFBGs with SNL do not admit a typical S-shaped hysteresis curve in their input-output characteristics at  $\delta = 0$ $cm^{-1}$.  A natural question that comes to mind is what happens to the OB (OM) curves for smaller values of the detuning parameter?, i.e., operating wavelengths of incident light close to the synchronous wavelength. We choose the positive values of the detuning parameter closer to the synchronous wavelength and investigate the nonlinear response of the proposed system via numerical simulations to address this query. 
	  
	  In Fig. \ref{fig3}(a) ($L = 20$ $cm$), we observe two distinct regions in the input-output characteristics curves. In region 1, increasing the input intensities induces sharp variations in the output intensities, leading to the ramp-like first stable states ($0<P_0<P_{th}^{r}$), as shown in Fig. \ref{fig3}(a).  As we tune the input intensity further, the system's output jumps to the second stable state branch. A bistable region with a narrow hysteresis width forms between the first and the second stable branch. The output intensities show gradual variations against the increasing input intensities ($P_{th}^{r}<P_0<P_{th}^{\uparrow}$). The output jumps to the next stable branch for ($P_0>P_{th}^{\uparrow}$). The output intensities vary gradually for $P_0>P_{th}^{\uparrow}$ at $L = 20$ $cm$. When the input intensities decrease, the system returns to the second stable branch at $P_{th}^{\downarrow}$.  Region 2 represents the values of input intensities for which the system's output is bistable ($P_{th}^{\downarrow}<P_0<P_{th}^{\uparrow}$).  Thus, the overall shape of the curves looks like a mix of ramp-like and S-shaped OB curves. In Fig. \ref{fig3}(a), the switching intensities of mixed OM curves are high. When the light incidence direction reverses, the switching intensities of mixed OM curves decrease dramatically, as shown in Fig. \ref{fig3}(b). An increase in the detuning parameter reduces the switch-up and down intensities of the mixed OM curves, as shown in Figs. \ref{fig3}(b) and (c). At the same time, an increase in the SNL parameter leads to an undesirable increase in the switching intensities, as shown in Fig. \ref{fig3}(d).

	\subsection{ S-shaped OB (OM) curves}
	\label{Sec:5A}
	\begin{figure*}
		\centering	
		\includegraphics[width=0.25\linewidth]{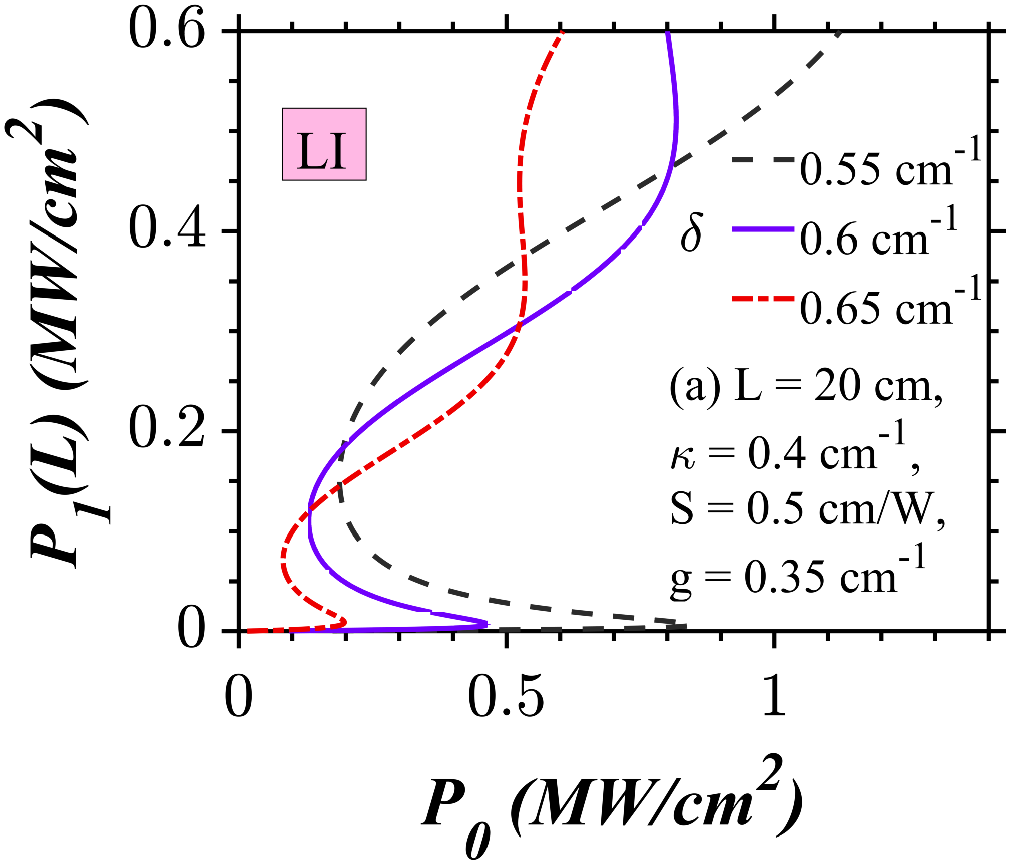}\includegraphics[width=0.25\linewidth]{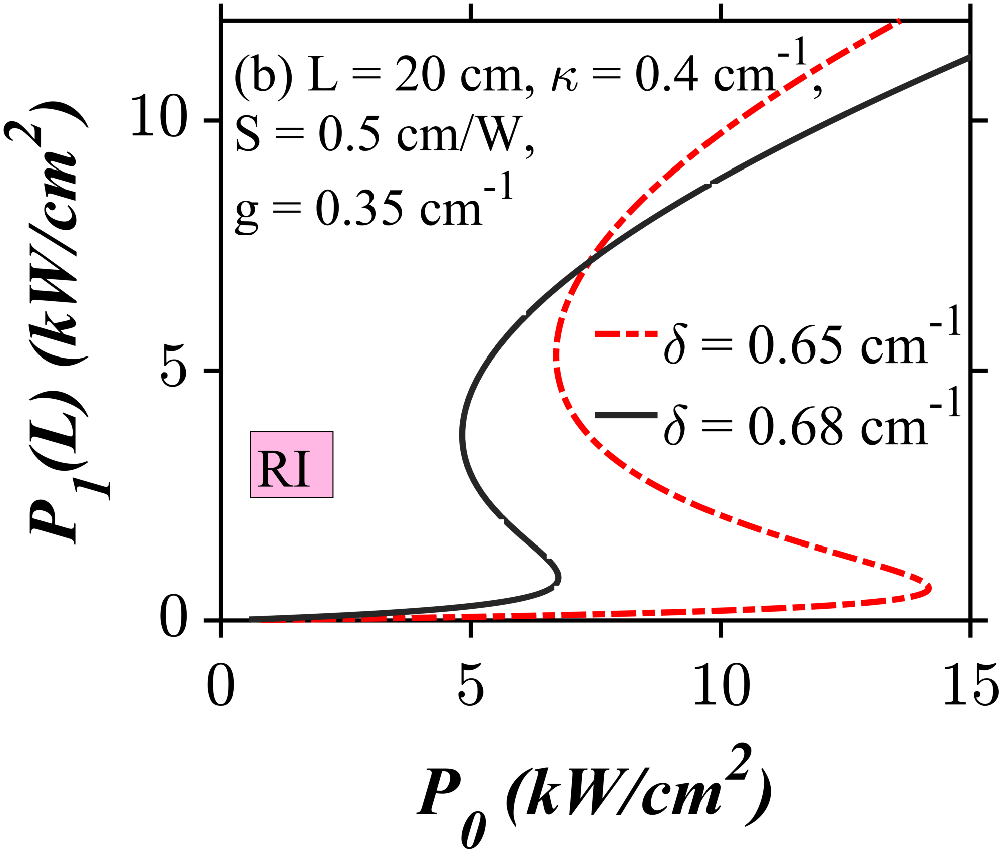}\includegraphics[width=0.25\linewidth]{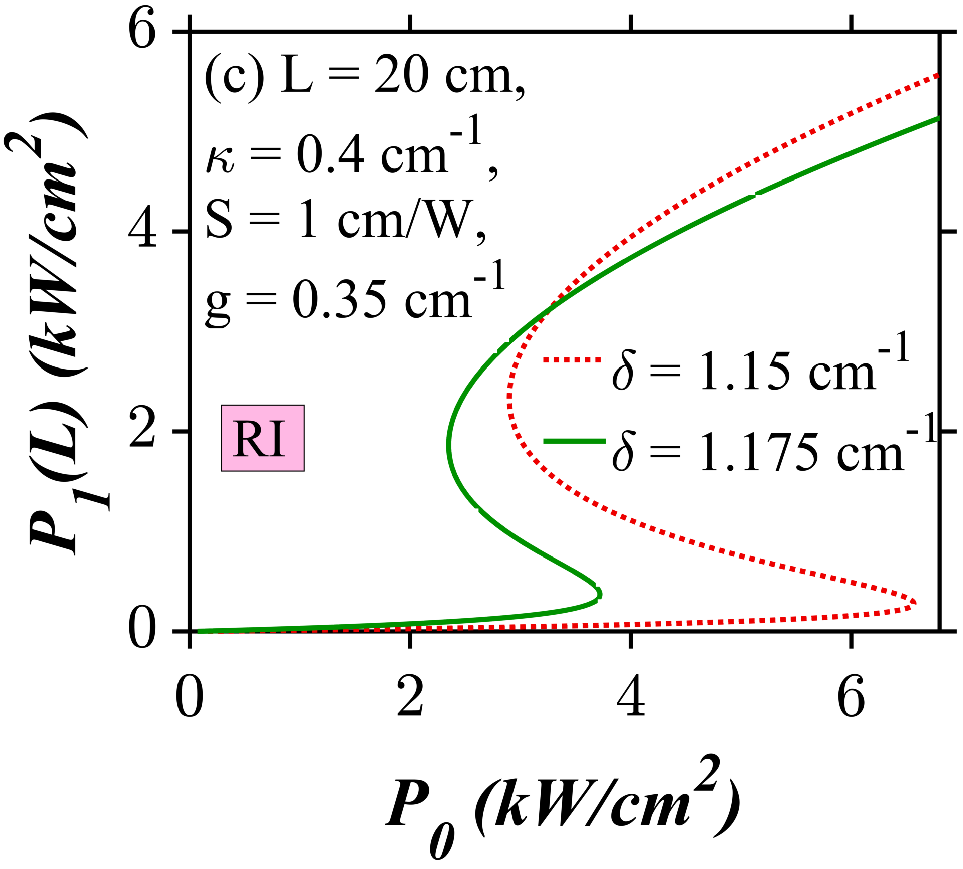}\includegraphics[width=0.25\linewidth]{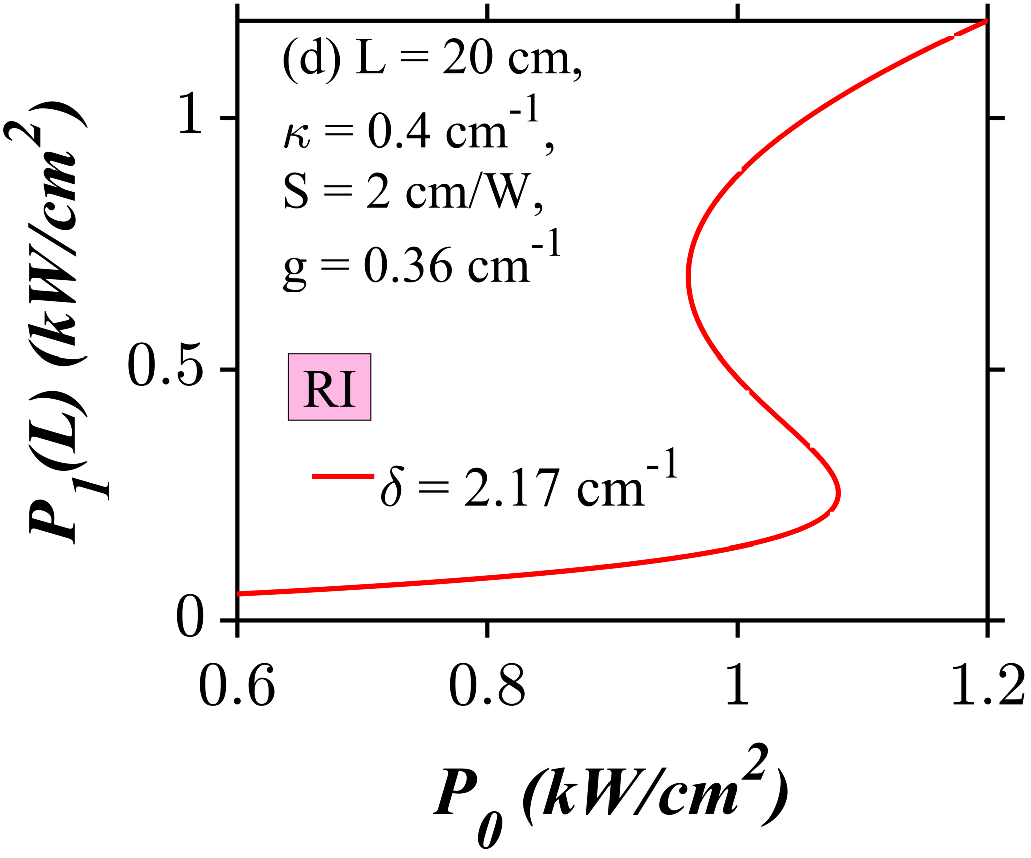}
		\\\includegraphics[width=0.25\linewidth]{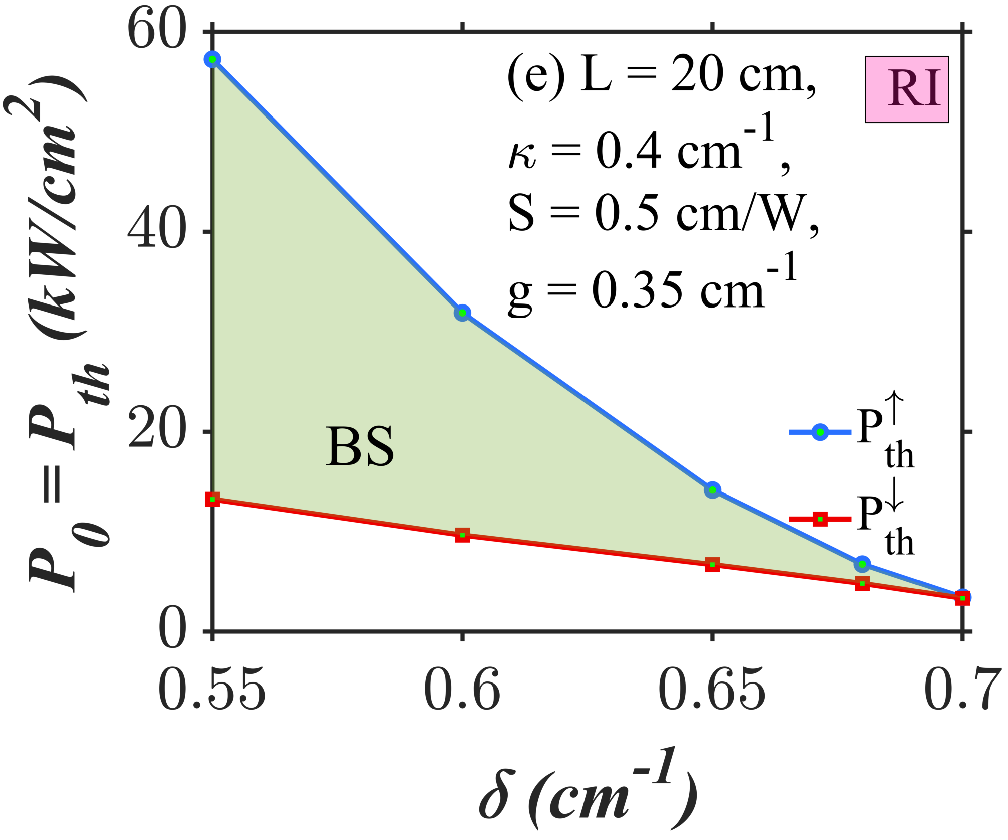}\includegraphics[width=0.25\linewidth]{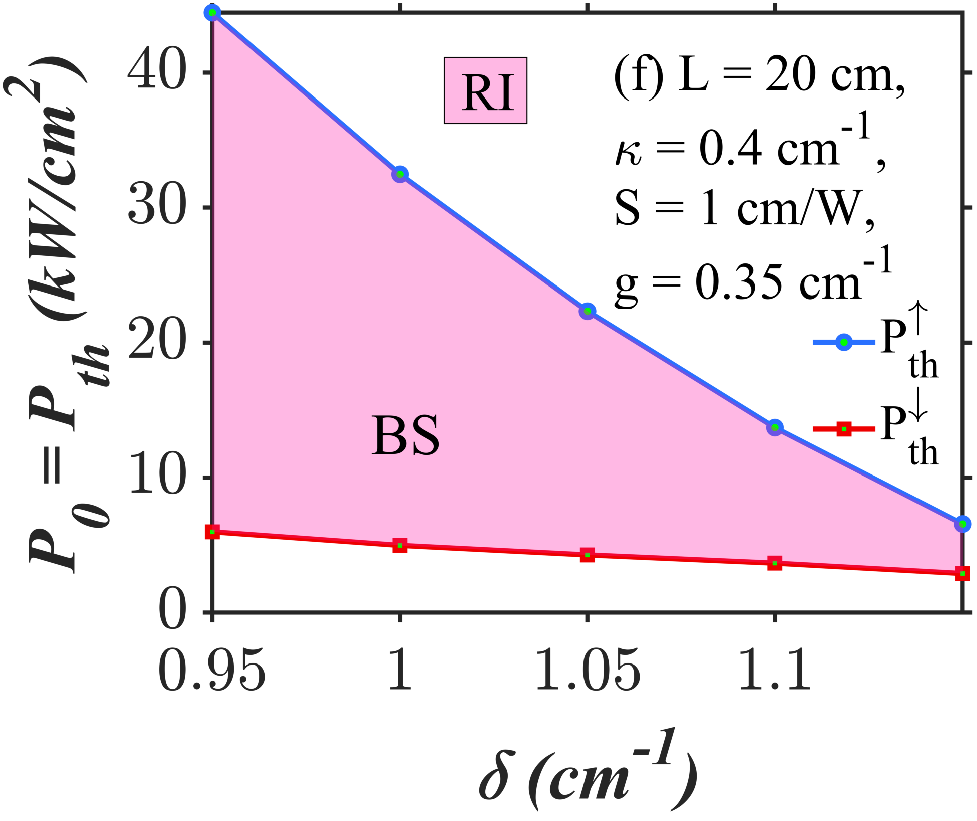}\includegraphics[width=0.25\linewidth]{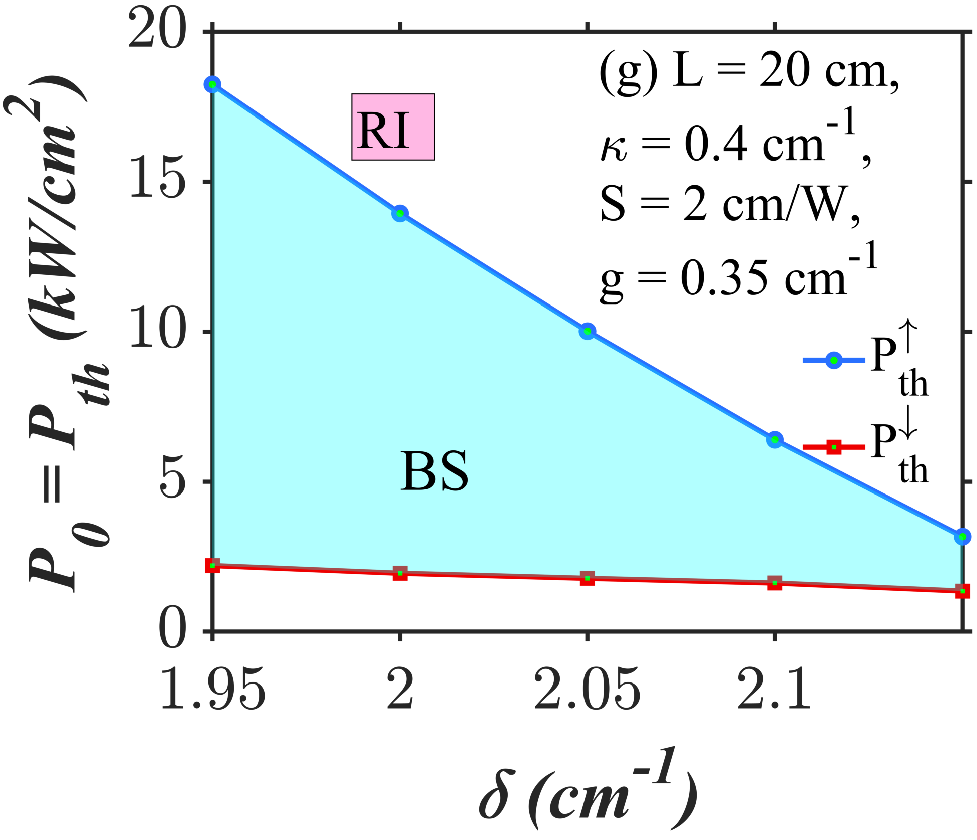}\includegraphics[width=0.25\linewidth]{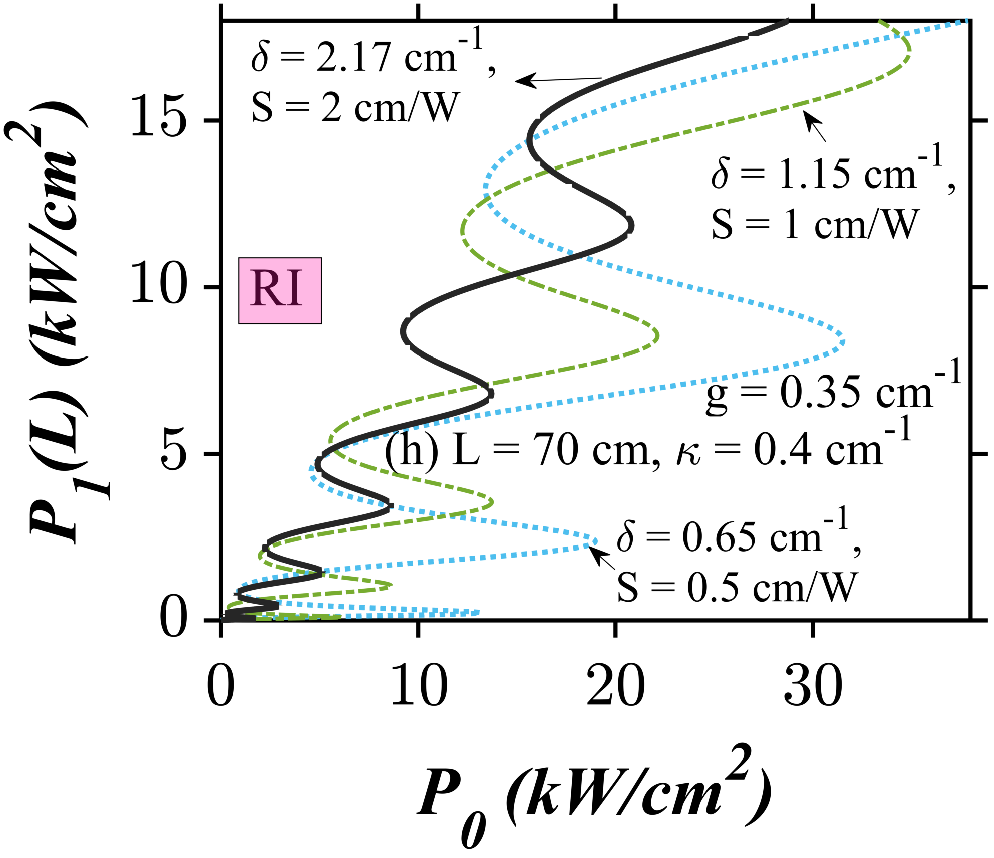}
		\caption{Frequency detuning induced ultralow power S-shaped OB curves (Figs. (a) - (g)) in an unbroken PTFBG with SNL  at $L = 20$ $cm$ and $\kappa = 0.4$ $cm^{-1}$. The direction of light incidence is (a) left and is right in the rest of the plots shown in the top panel.  (e) -- (g) Variations in switching intensities and hysteresis width against the detuning and SNL.  (h) Low-power S-shaped OM curves for a higher length at $L = 70$ $cm$. }
	\label{fig4}
	\end{figure*}
	
	The search for realizing S-shaped OB (OM) curves motivated us to study the switching dynamics of an unbroken PTFBG with SNL for high values of the detuning parameter. The system offers control over the switch-up and down intensities ($P_{th}^{\uparrow}$ and $P_{th}^{\downarrow}$ ) via an independent tuning of one or more system parameters, as shown in Figs. \ref{fig4}(a) -- (d). In the first approach, they get reduced via the frequency detuning, as shown in Figs. \ref{fig4}(a) -- (c). The higher the difference between the operating and Bragg wavelengths, the lesser is their values, and the narrower is the hysteresis width. In the second approach, the switching intensities decrease under the refractive index condition, as shown in Fig. \ref{fig4}(b).
	
	 Tuning the SNL parameter to a high value ($S \ge 1$ $cm/W$) can perform the equivalent job of reducing the switch-up and down intensities at fixed values of the gain and loss parameter, as shown in Fig. \ref{fig4}(c).   An alternate solution to reduce the switch-up and down intensities is to increase the gain and loss levels (from $g = 0.35$ to 0.36 $cm^{-1}$) at a fixed value of the detuning parameter ($\delta$). In this fashion, they reduce to a level of $<$ 1.1 $kW/cm^2$, provided that the value of SNL is high ($S = 2$ $cm/W$), as portrayed by Fig. \ref{fig4}(d). 
 
 In equivalent normalized units, the curve indicated by the solid line in Fig. \ref{fig4}(d) features switching intensities less than 0.0011. In the literature, the lowest-ever switching intensities recorded so far in the context of PTFBGs lie in the range $0.04<P_0<0.05$ (theoretically) \cite{sudhakar2022inhomogeneous, sudhakar2022low}. However, the curve indicated by the solid line in Fig. \ref{fig4}(d) features switching intensities less than 0.0011, significantly lesser the lowest-ever recorded value in the literature.
	
	How does reversing the direction of light incidence lower the switching intensities dramatically ? There seem to exist a few scientific articles that address this natural query. When the light gets launched from the other input surface of the PTFBG, the two counter-propagating modes interact constructively in the gain regions \cite{kulishov2005nonreciprocal}. The asymmetric nature of the complex refractive index distribution prohibits a constructive interaction between the modes for the left light incidence condition. Even though this interactive picture traces its origin to the linear domain, it applies to nonlinear PTFBGs as well. Studies on the spatial distribution of the total electric field (superposition of the forward and backward traveling waves) suggest that the maxima of the optical field lie in the gain region for the right light incident conditions and vice-versa \cite{komissarova2019pt}. The nonreciprocal switching at ultra-low power intensities arises from such a mutual arrangement of the optical field maxima in the gain-loss structure. 
	
	The S-shaped OB curves occur for a set of specific values of the detuning parameter ($\delta_{min} < \delta < \delta_{max}$) that vary with the value of SNL, as depicted by the parametric plots in Figs. \ref{fig4}(e) -- (g).  Note that the hysteresis width and switching intensities are the lowest and highest at $\delta_{max}$ and $\delta_{min}$, respectively.
	
 Ultralow-power S-shaped OB shown in the top panel of Fig. \ref{fig4} transforms into low-power S-shaped OM curves upon  increasing the device length to $L = 70$ $cm$, as shown in Fig. \ref{fig4}(h). The interplay among the detuning, SNL, gain, and loss parameters reduces the switch-up and down intensities of the S-shaped OM curves. In short, the individual roles of different system parameters on the switching intensities remain the same irrespective of the value of the device length. It is worth noting that the model presented here is not limited to FBGs but applies to periodic devices like photonic crystals that share similarities in bandgap structure and operation with FBGs.
 
 \section{OB and OM in the broken $\mathcal{PT}$-symmetric regime: $L = 20$ $cm$}\label{Sec:5}

 \begin{figure}
 	\centering	\includegraphics[width=0.5\linewidth]{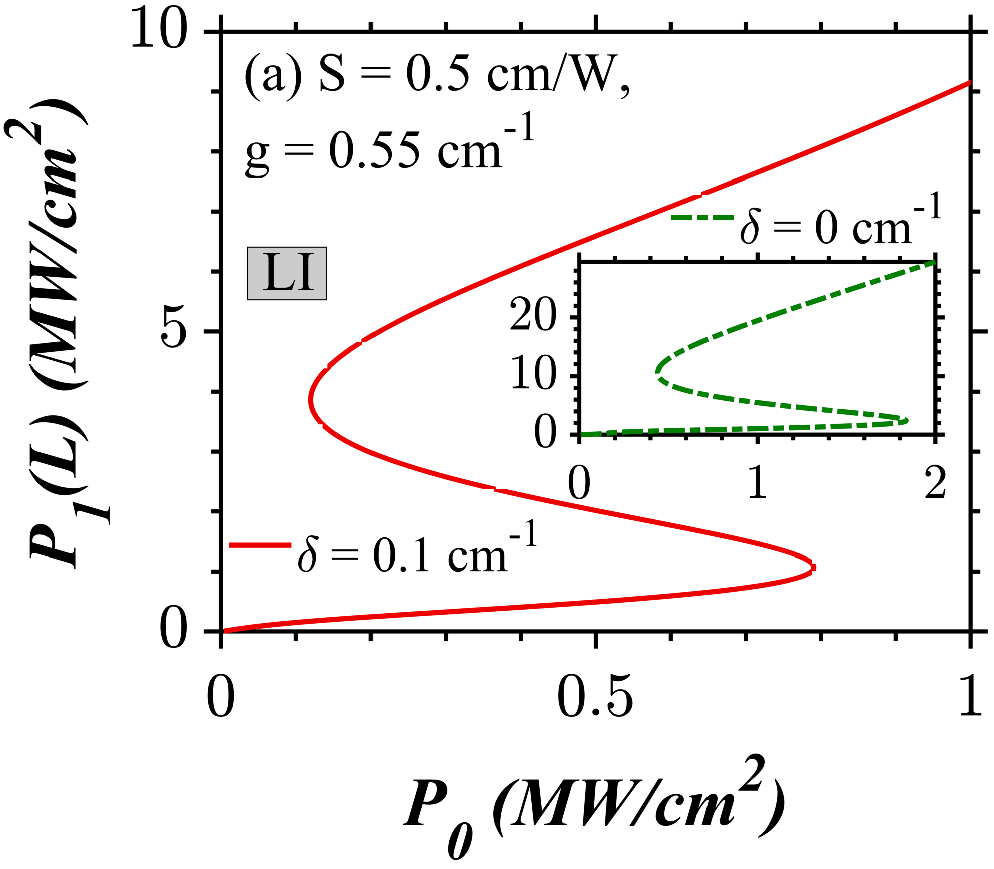}\includegraphics[width=0.5\linewidth]{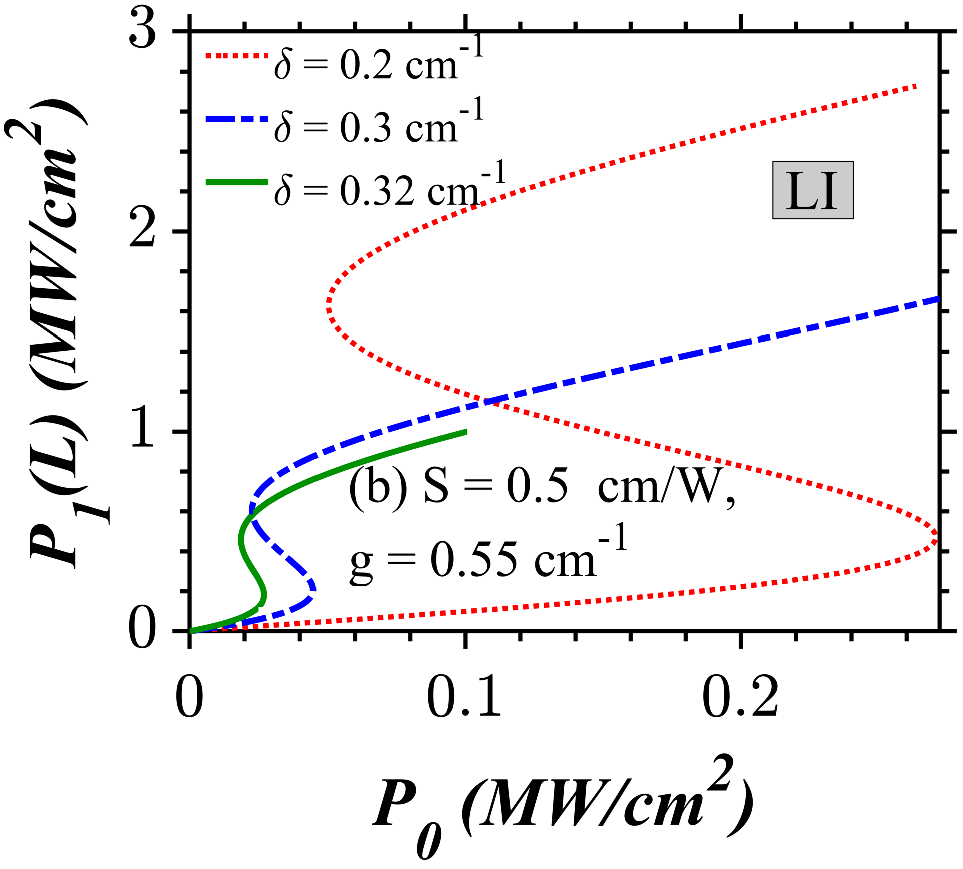}\\\includegraphics[width=0.5\linewidth]{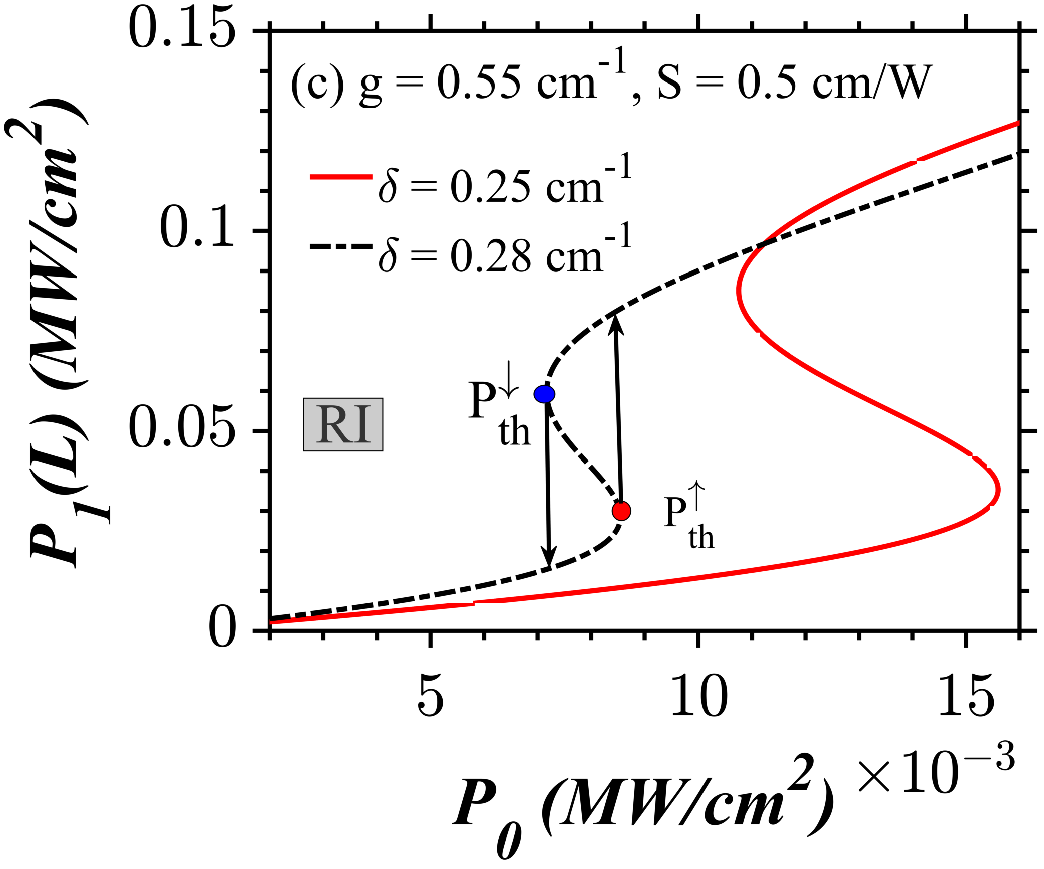}\includegraphics[width=0.5\linewidth]{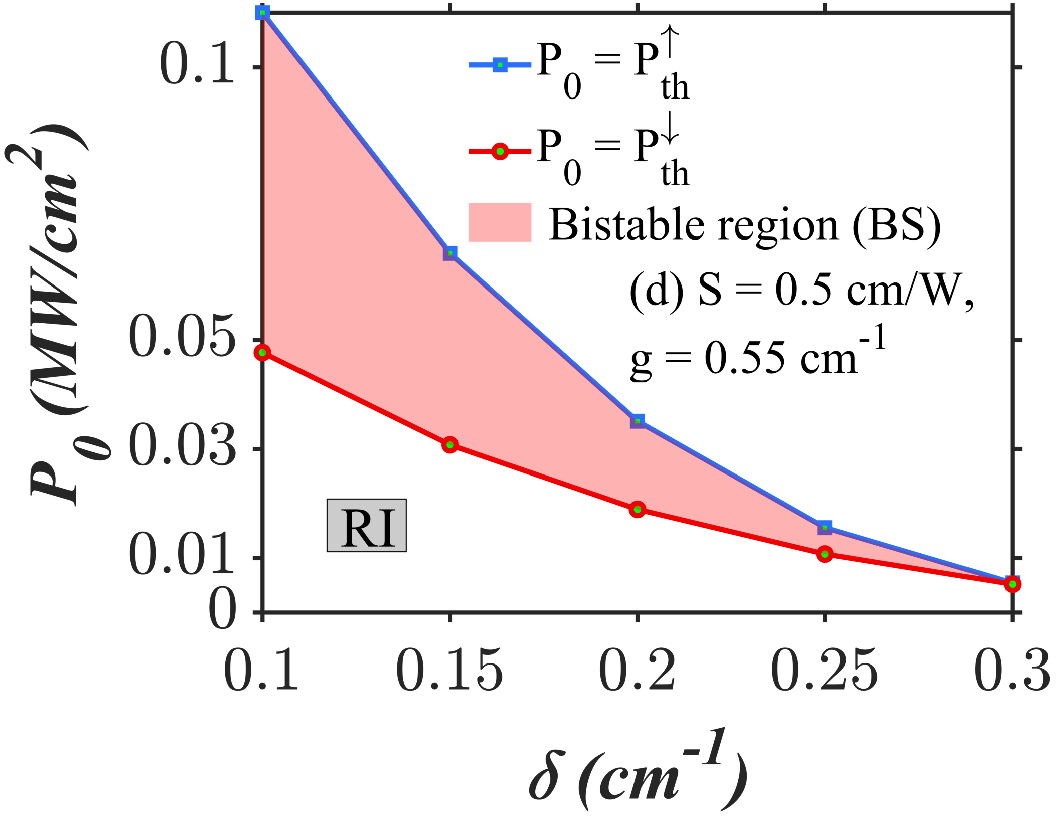}\\\includegraphics[width=0.5\linewidth]{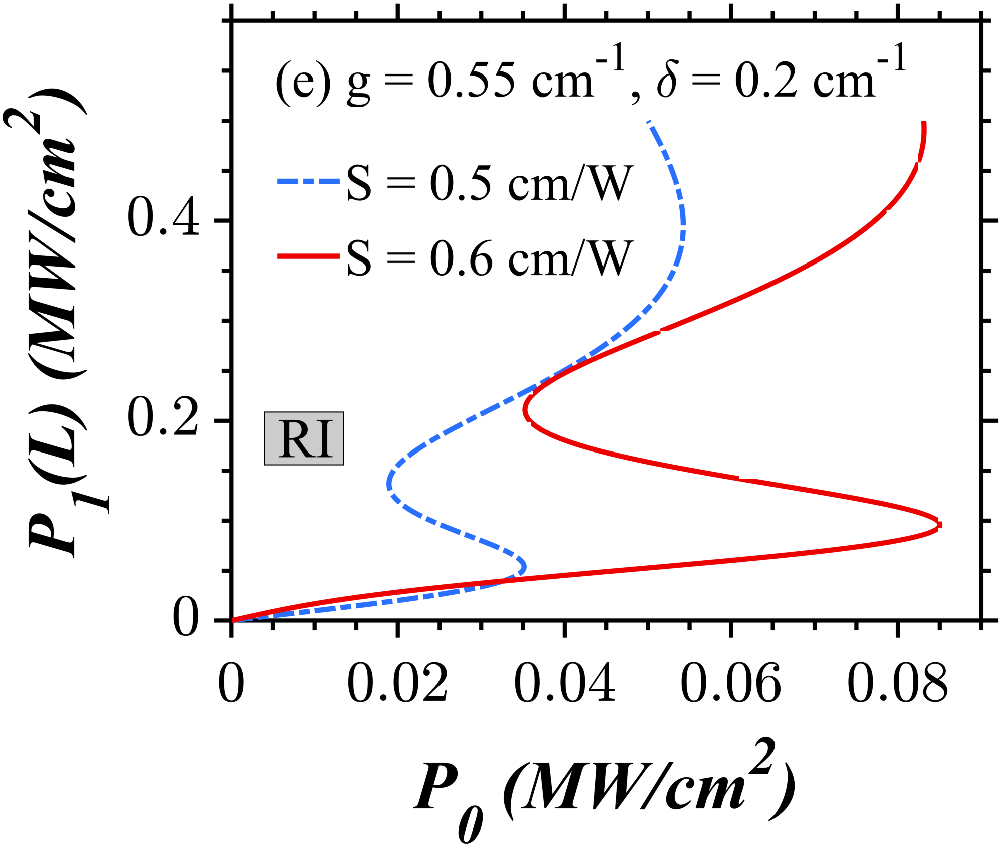}\includegraphics[width=0.5\linewidth]{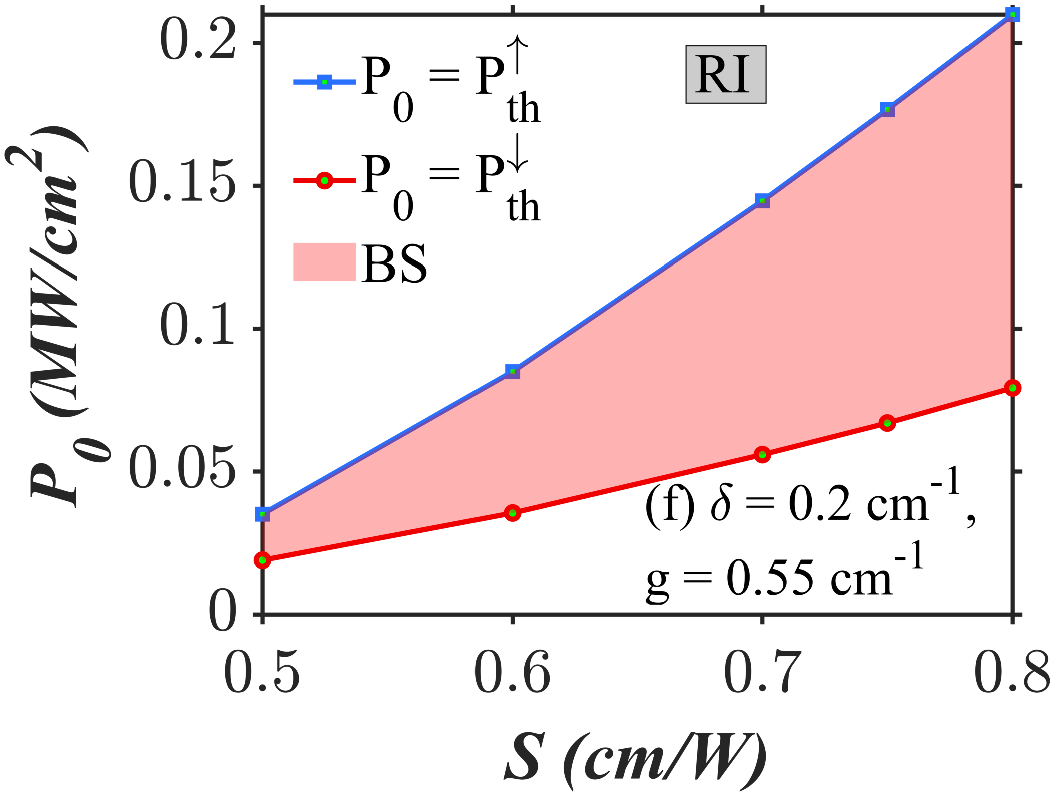}
 	\caption{ S-shaped OB curves generated by PTFBGs with SNL operating in the broken $\mathcal{PT}$- symmetric regime. The light incidence direction is  left in (a, b), and is right in (c-f). Middle and bottom panels: Role of detuning and SNL parameters on the switching intensities.}
 	\label{fig5}
 \end{figure}
 
 \begin{figure}
 	\centering	\includegraphics[width=0.5\linewidth]{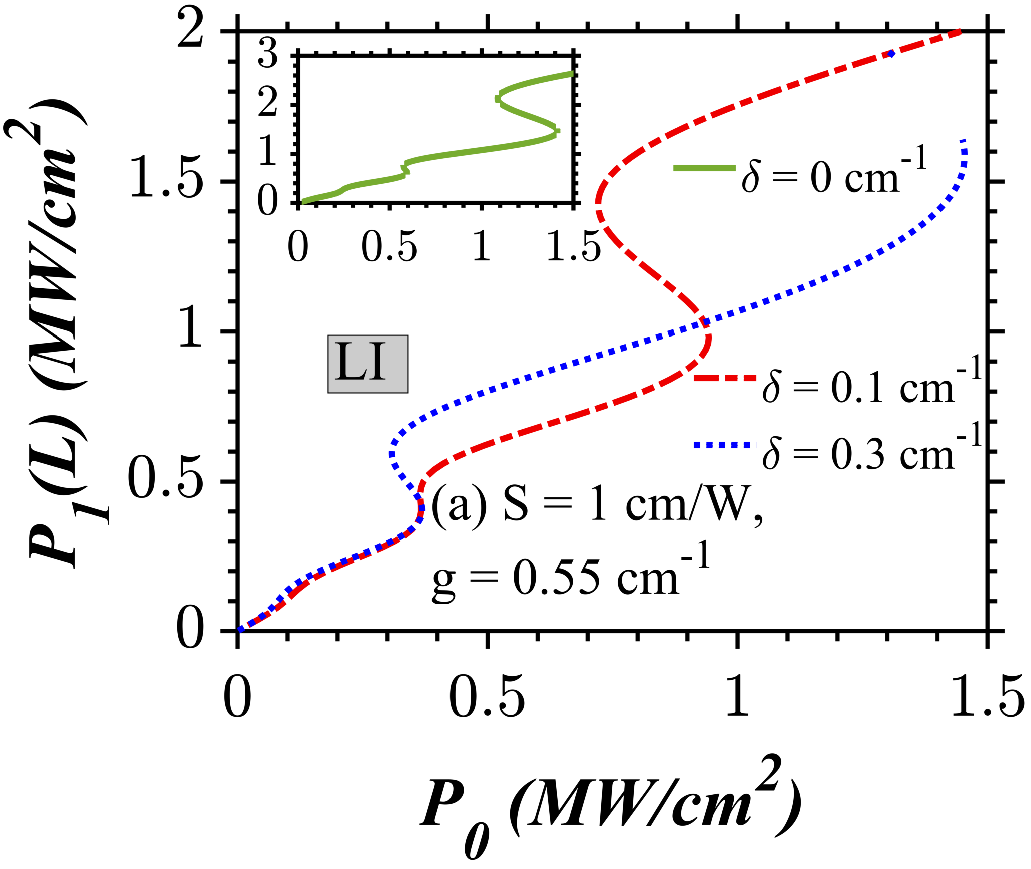}\includegraphics[width=0.5\linewidth]{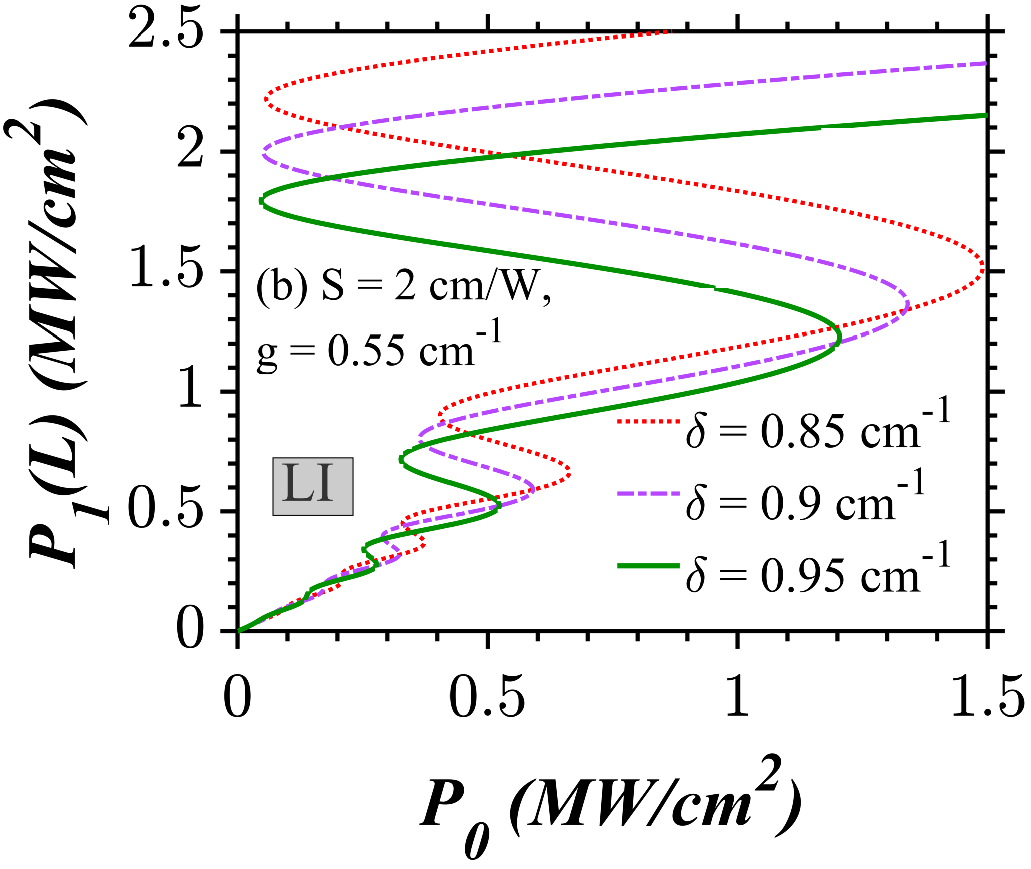}\\\includegraphics[width=0.5\linewidth]{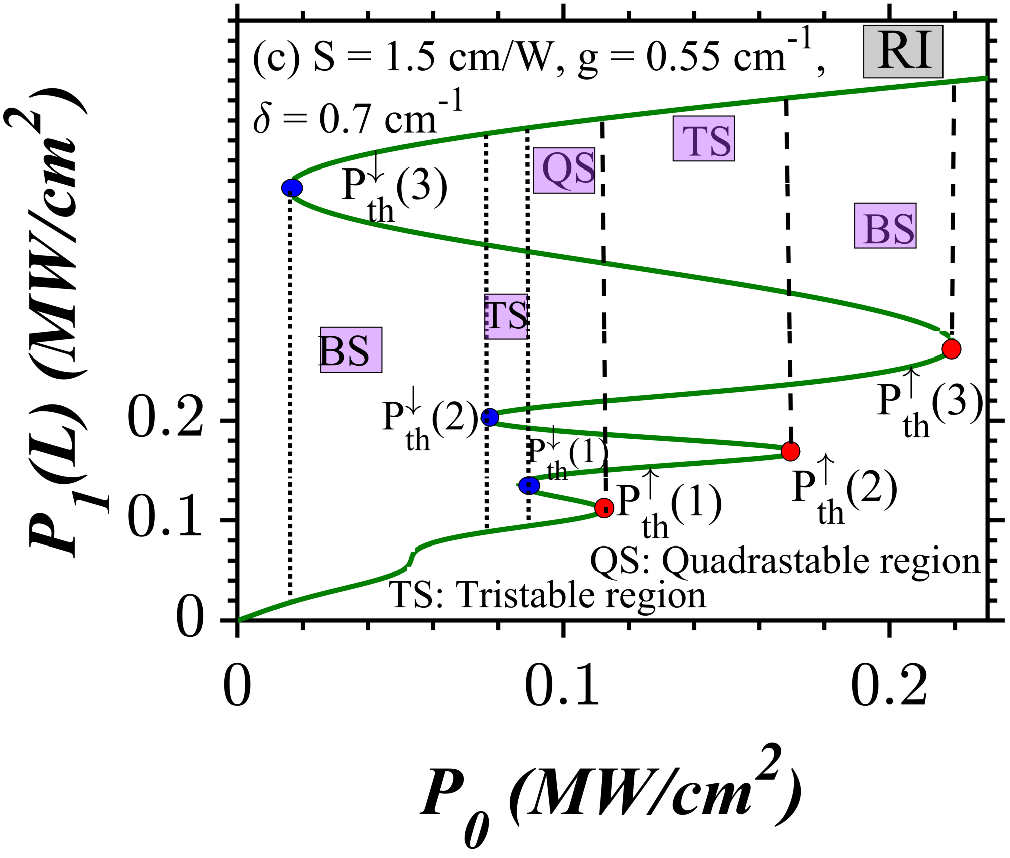}\includegraphics[width=0.5\linewidth]{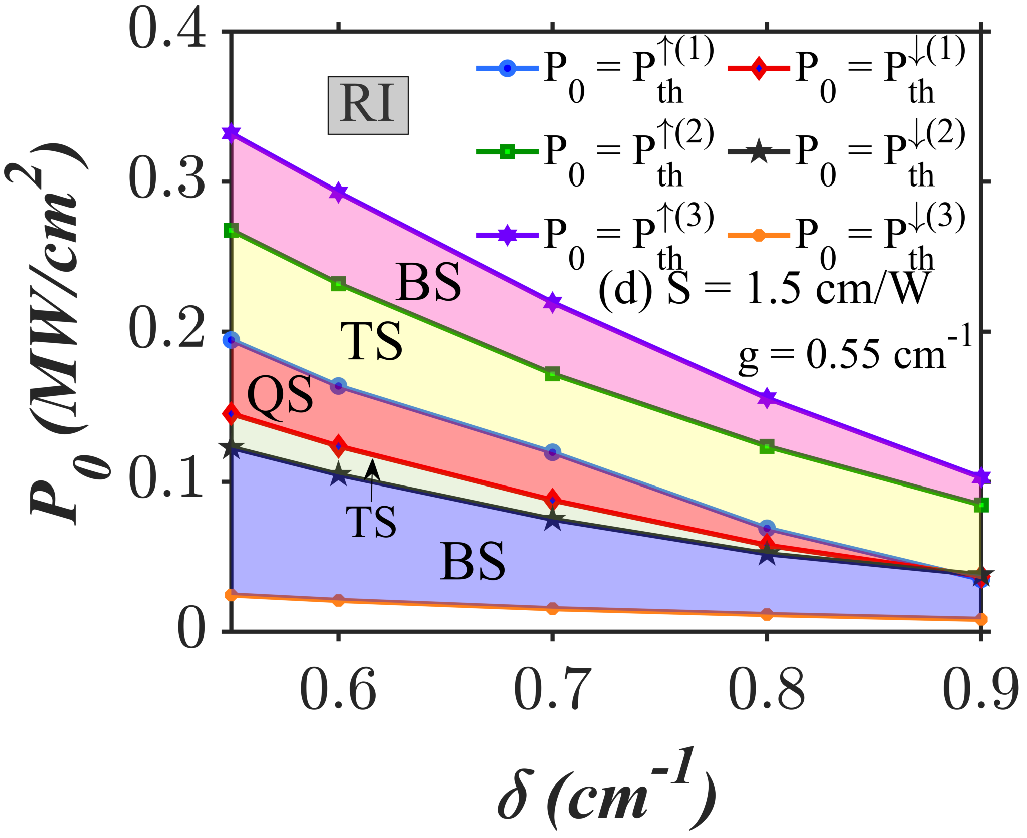}\\\includegraphics[width=0.5\linewidth]{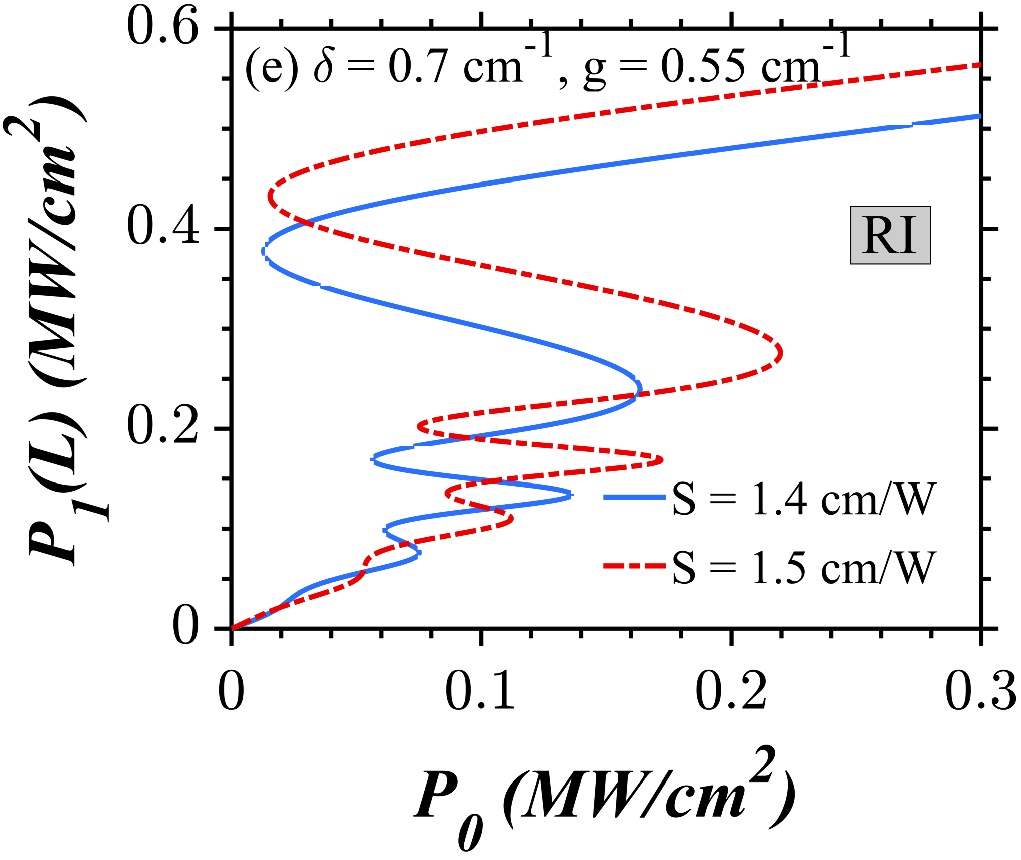}\includegraphics[width=0.5\linewidth]{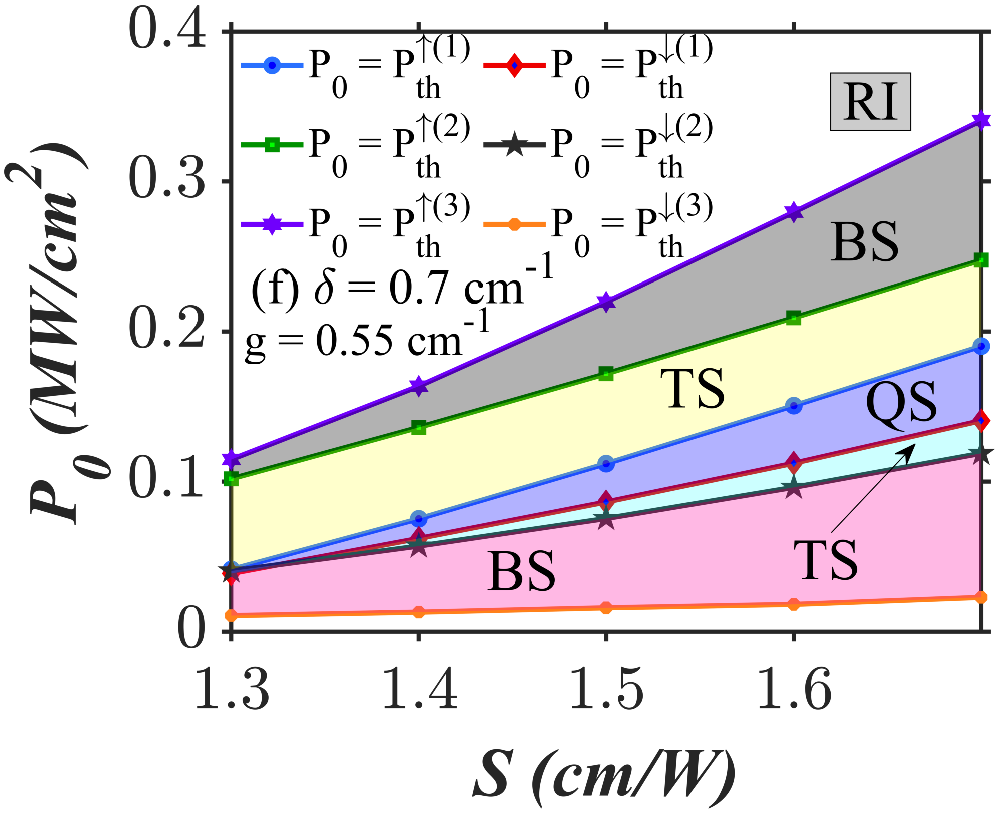}
 	\caption{ Ramp-like OB and OM curves generated by broken PTFBGs with SNL. The direction of light incidence is  left in (a, b), and is right in (c-f). Middle and bottom panels: Role of detuning and SNL parameters on the switching intensities. Also, (c) depicts various stable regions, namely, bistable (BS), tristable (TS), and quadrastable (QS) in a ramp-like OM curve.   }
 	\label{fig6}
 \end{figure}
 \begin{figure*}
 	\centering	\includegraphics[width=0.3\linewidth]{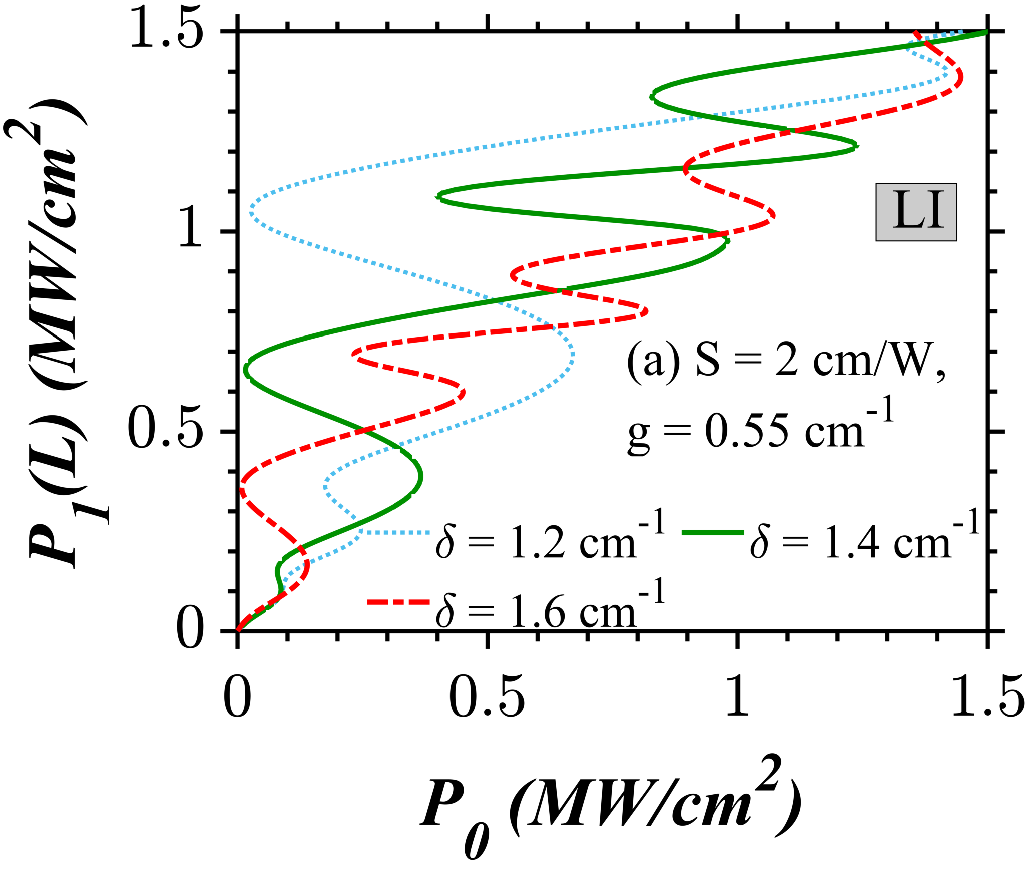}\includegraphics[width=0.3\linewidth]{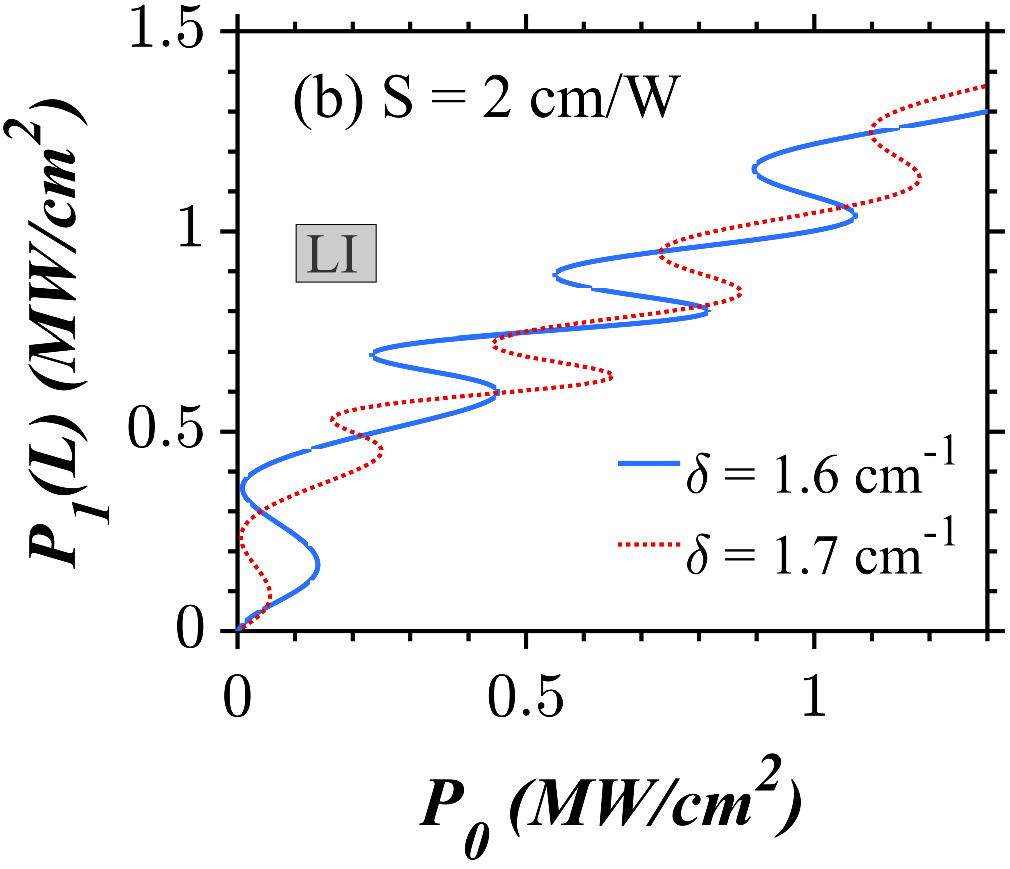}\includegraphics[width=0.3\linewidth]{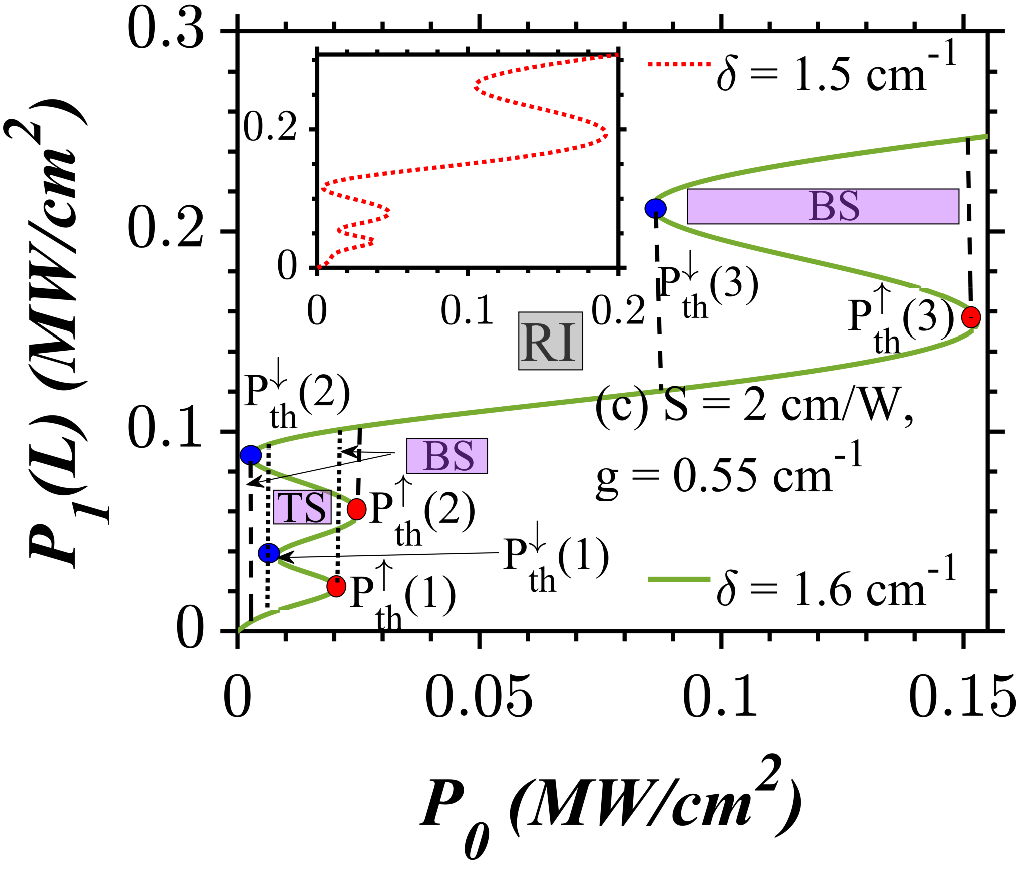}\\\includegraphics[width=0.3\linewidth]{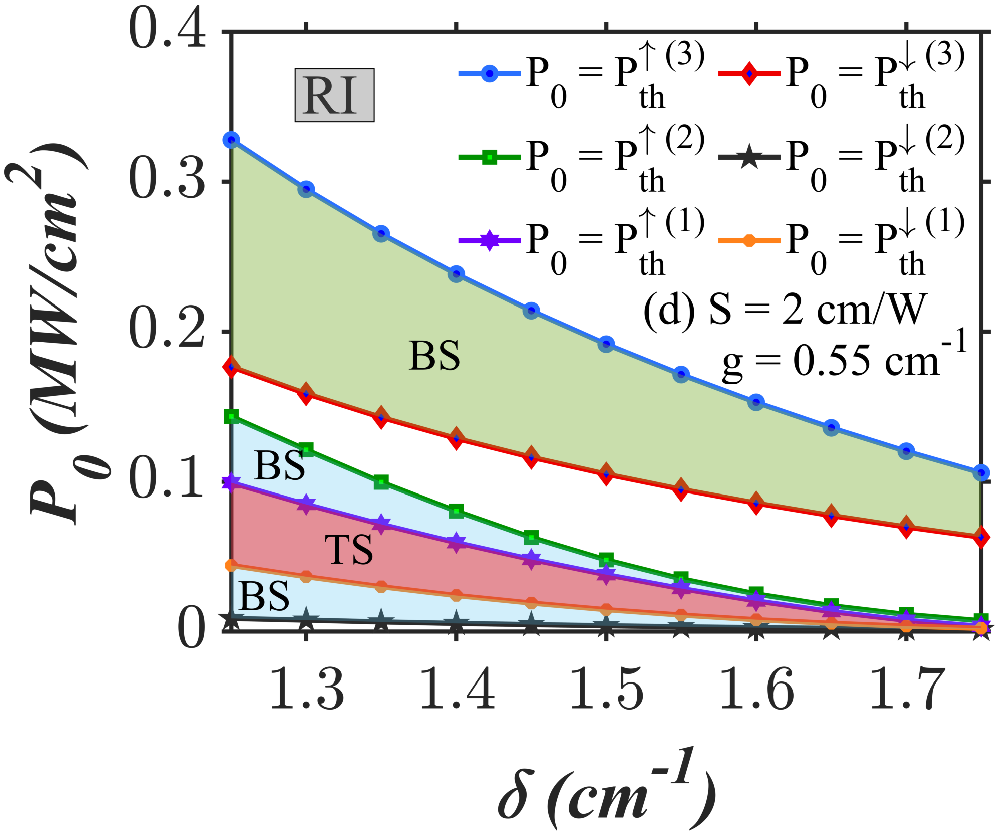}\includegraphics[width=0.3\linewidth]{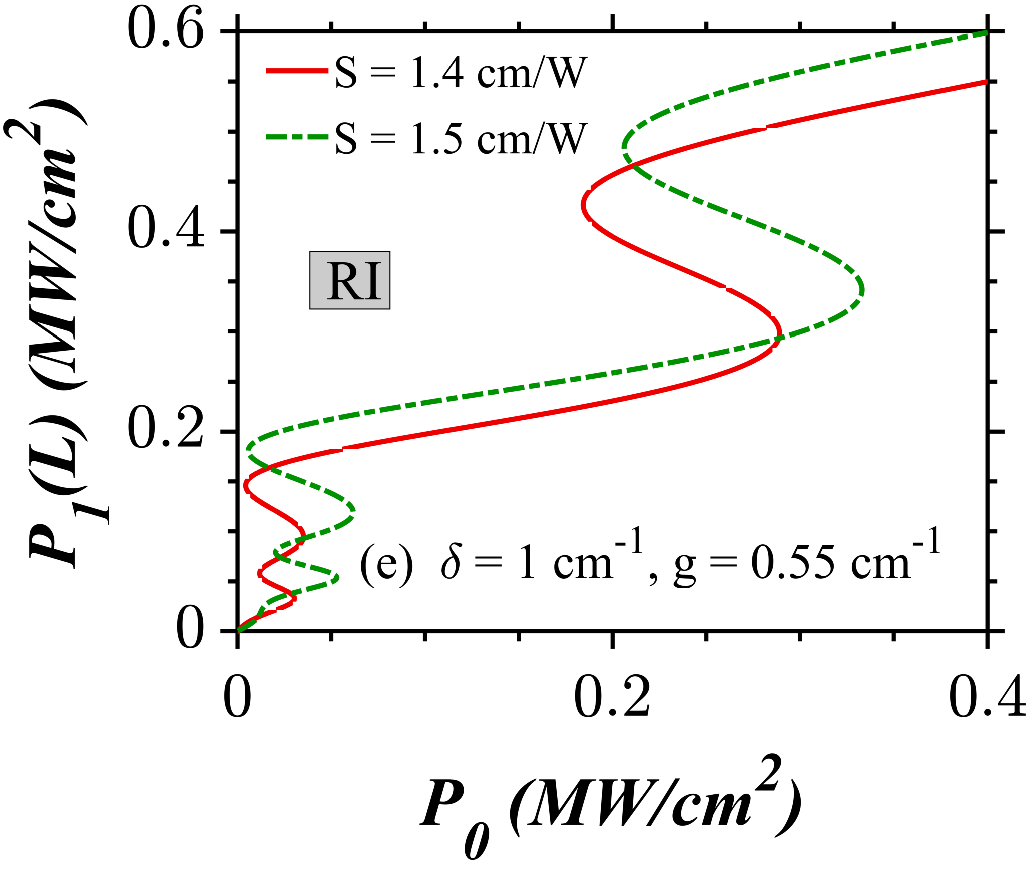}\includegraphics[width=0.3\linewidth]{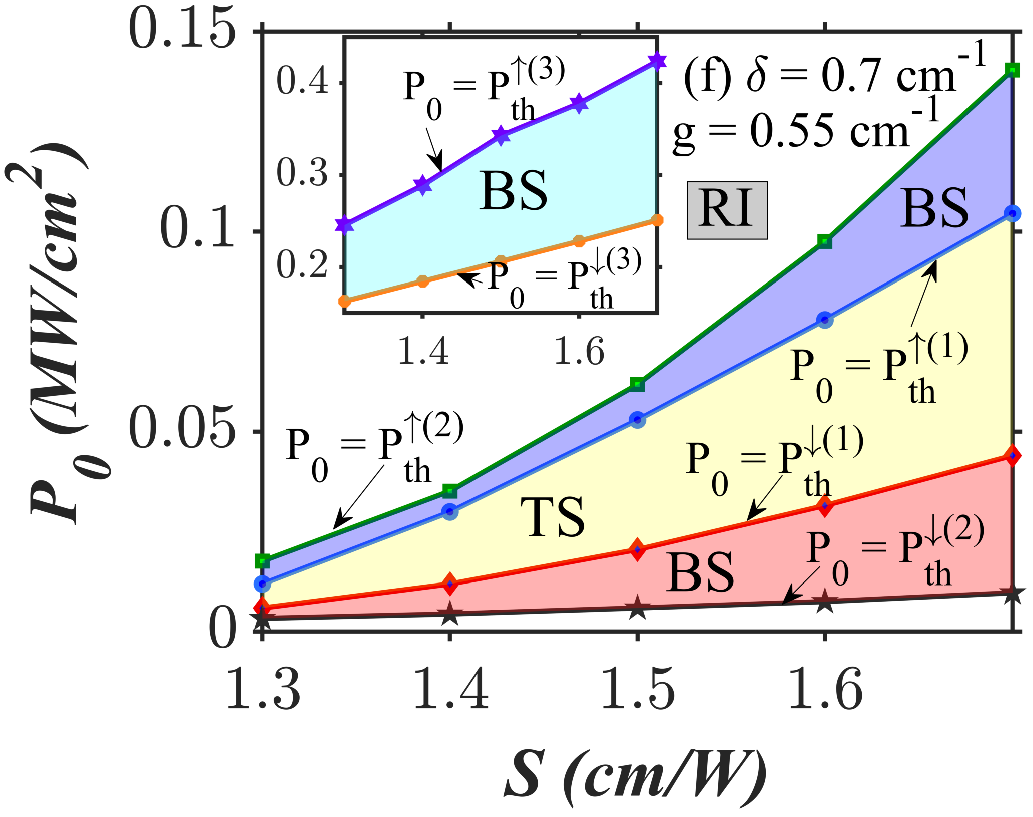}
 	\caption{Plots depicting the same dynamics as in Fig. \ref{fig6} with different values of detuning and SNL parameters. The input-output characteristics curves feature mixed OM curves due to the increase in the value of SNL parameter. }
 	\label{fig7}
 \end{figure*}
 
 \subsection{S-shaped OB curves  at $L = 20$ cm}\label{Sec:6A}
 The system generates S-shaped OB curves in the broken $\mathcal{PT}$- symmetric regime for lower values of saturable nonlinearity, as shown in Figs. \ref{fig5}(a) and (b). In PTFBGs operating in the broken $\mathcal{PT}$-symmetric regime, the S-shaped OB curves are uncommon, deviating from the expected ramp-like OB and OM reported in the literature \cite{raja2019multifaceted,PhysRevA.100.053806,sudhakar2022inhomogeneous,sudhakar2022low}. The inclusion of SNL ensures that the PTFBG system produces ramp-like input-output characteristics in the unbroken $\mathcal{PT}$-symmetric regime, as seen in Secs. \ref{Sec:3}, and S-shaped OB/OM curves in the broken $\mathcal{PT}$-symmetric regime, as depicted in Fig. \ref{fig5}. This behavior contrasts with the input-output characteristics of PTFBGs without SNL, which exhibit S-shaped OM curves in the unbroken regime and ramp-like OM curves in the broken regime \cite{raja2019multifaceted,PhysRevA.100.053806,sudhakar2022inhomogeneous,sudhakar2022low}.

 Figure \ref{fig5}(c) reveals that reversing the direction of light incidence decreases the switching intensities of the S-shaped OB curves. The observed result directly arises from the maxima of the total optical field naturally aligning within the gain regions under the right light incidence condition \cite{kulishov2005nonreciprocal}, a characteristic shared among all examined nonlinear PTFBG configurations to date \cite{komissarova2019pt,raja2019multifaceted,PhysRevA.100.053806}. The switching intensities and the hysteresis width decrease with an increase in the detuning parameter for fixed values of SNL, as shown in Fig. \ref{fig5}(d). On the other hand, if the value of the SNL parameter increases at fixed values of the detuning parameter, it produces an undesirable effect in the form of an increase in the switching intensities, as shown in Figs. \ref{fig5}(e) and (f). When the detuning parameter is very high, the OB curves disappear. The disappearance of OB curves under this condition does not imply a complete elimination of the system's potential for further manipulating the switching dynamics, as FBGs offer multiple avenues for altering the steering dynamics. However, it is essential to carefully adjust the other system parameters in numerical simulations for optimal results, as detailed in the following sections.

 \subsection{Ramp-like OB and OM curves at $L = 20$ cm}\label{Sec:6B}
 As the nonlinearity value increases (above $S = 0.8$ $cm/W$), a transformation from S-shaped OB to ramp-like OB occurs, as confirmed by the plots in Fig. \ref{fig6}(a). When the nonlinearity value is further increased (at fixed input intensities), the input-output characteristics curve generates a ramp-like OM curve, as depicted in Fig. \ref{fig6}(b). The width of the successive hysteresis curves emerging above the initial ramp-like stable branch expands with the tuning of input intensity. A dramatic reduction in switching intensities occurs under the reversal in the direction of light incidence, as shown in Fig. \ref{fig6}(c). With the frequency detuning, the switching intensities needed to transition from the first ramp-like stable state to the second decrease significantly, as illustrated in Fig. \ref{fig6} (d). Increasing the value of SNL parameter has an undesirable effect on the ramp-like OM curve, leading to higher switching intensities, as shown in Figs. \ref{fig6}(e) and (f). The system exhibits tristable (TS) and quadrastable (QS) outputs, representing three and four output states, respectively, for specific input intensity ranges, as shown in Figs. \ref{fig6}(d) and (f). Specifically, TS occurs when $P_{th}^{\downarrow(2)} \le P_0 < P_{th}^{\downarrow(1)}$ and $P_{th}^{\uparrow(1)} \le P_0 < P_{th}^{\uparrow(2)}$, while QS occurs in the range $P_{th}^{\downarrow(1)} \le P_0 < P_{th}^{\uparrow(1)}$. For other input intensities, the system remains bistable.

 \subsection{Low power mixed OM curves at $L = 20$ cm}\label{Sec:6C}	
 The system exhibits mixed OM curves for larger detuning parameter values, as depicted in Figs. \ref{fig7}(a) and (b). The OB curves take on a ramp-like shape at low intensities, transitioning to an S-shaped curve at higher intensities. Notably, mixed OM curves in Fig. \ref{fig7}(c) occur at lower intensities than those in Figs. \ref{fig7}(a) and (b), thanks to the reversal in the direction of light incidence., as depicted in Figs. \ref{fig7}(a) and (b).
  Similar to other types of OB and OM curves, the switching intensities associated with mixed OM curves decrease with an increase in the detuning parameter and increase with an increase in the SNL parameter, as illustrated in Figs. \ref{fig7}(d) -- (f). The system is bistable everywhere, except when the input intensity falls in the range $P_{th}^{\downarrow(1)}\le P_0 \le P_{th}^{\uparrow(1)}$, where the system supports tristability, as confirmed by Figs. \ref{fig7}(d) and (f). Thus, we can generate different OM curves like S-shaped, ramp-like, and mixed OM in the same system by tuning the values of SNL and detuning parameters.

 \subsection{Unusual OM curves at $L = 70$ cm}
 \label{Sec:6D}

 \begin{figure*}
 	\centering	\includegraphics[width=0.25\linewidth]{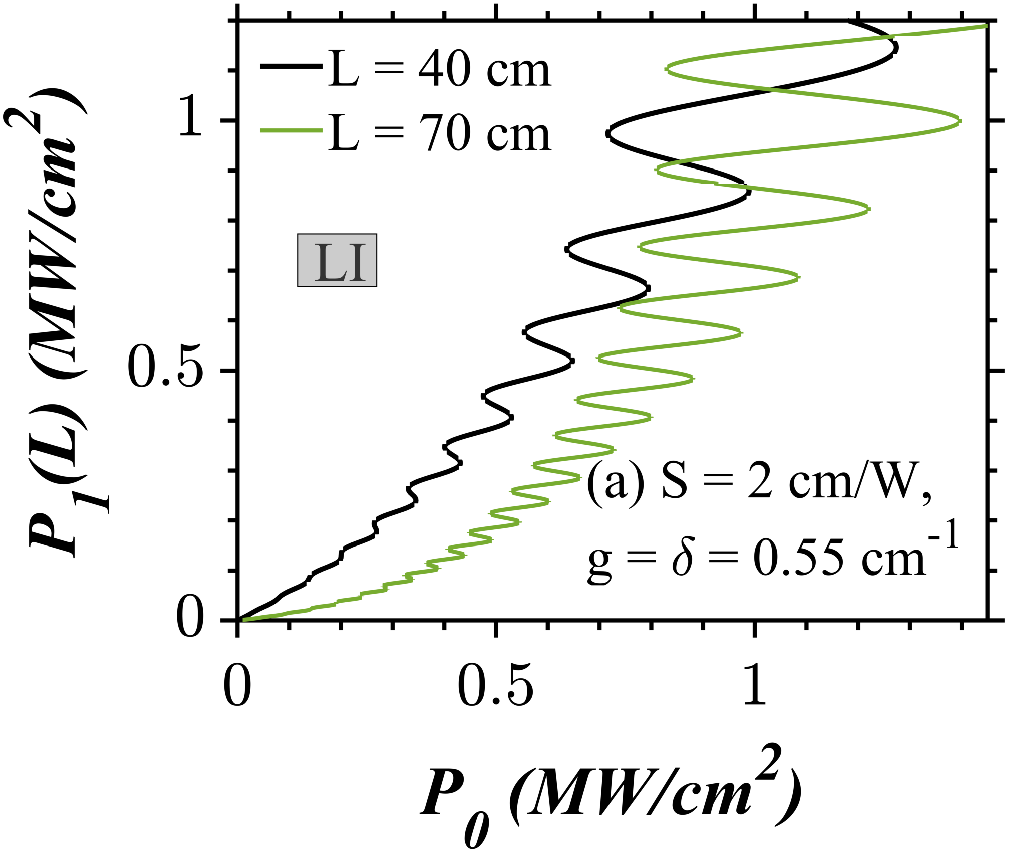}\includegraphics[width=0.25\linewidth]{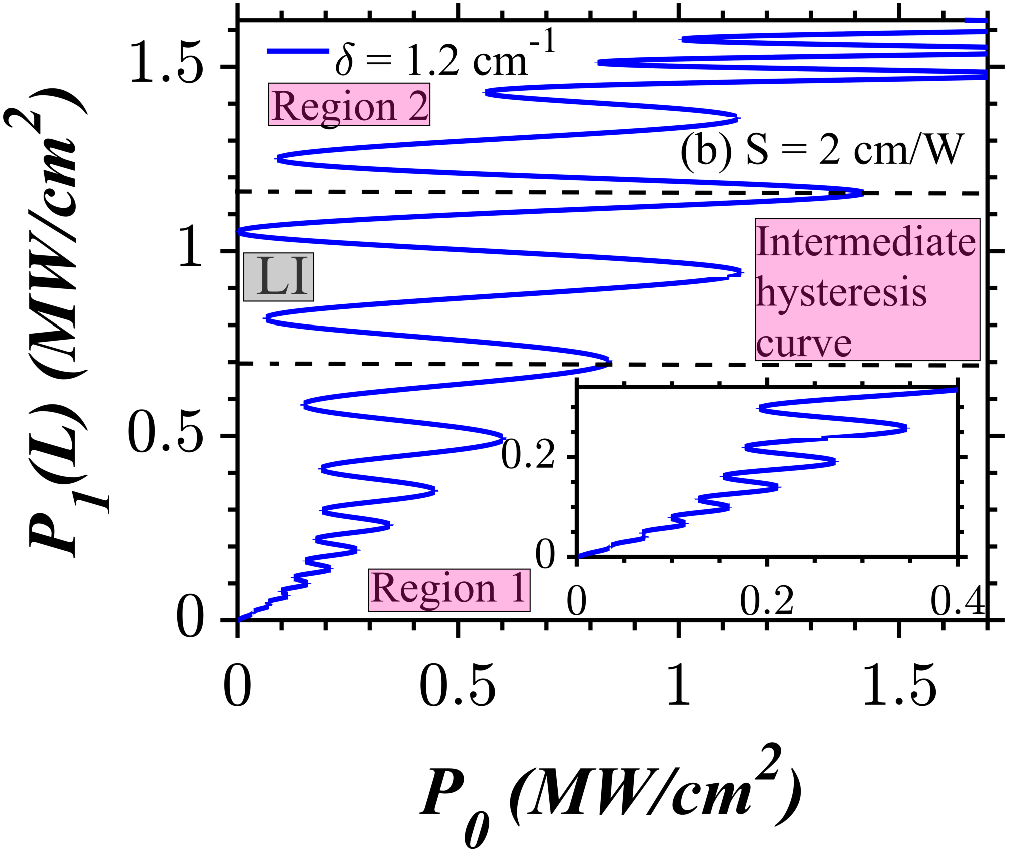}\includegraphics[width=0.25\linewidth]{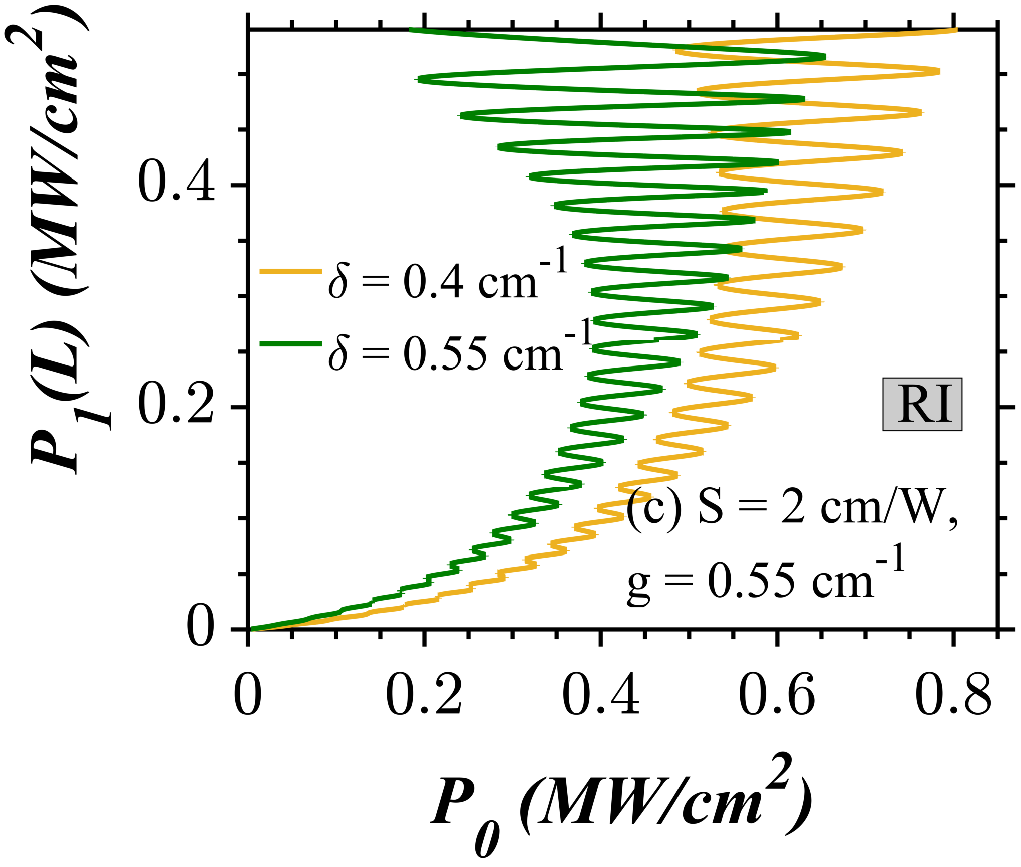}\includegraphics[width=0.25\linewidth]{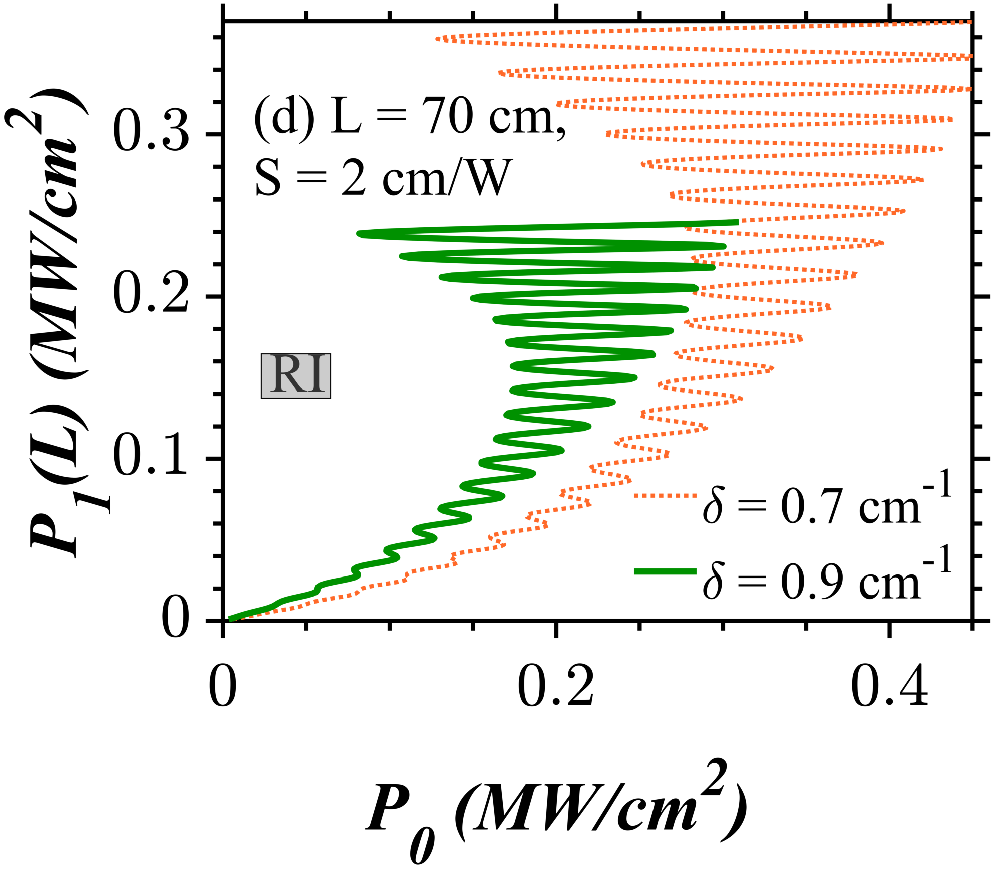}\\\includegraphics[width=0.25\linewidth]{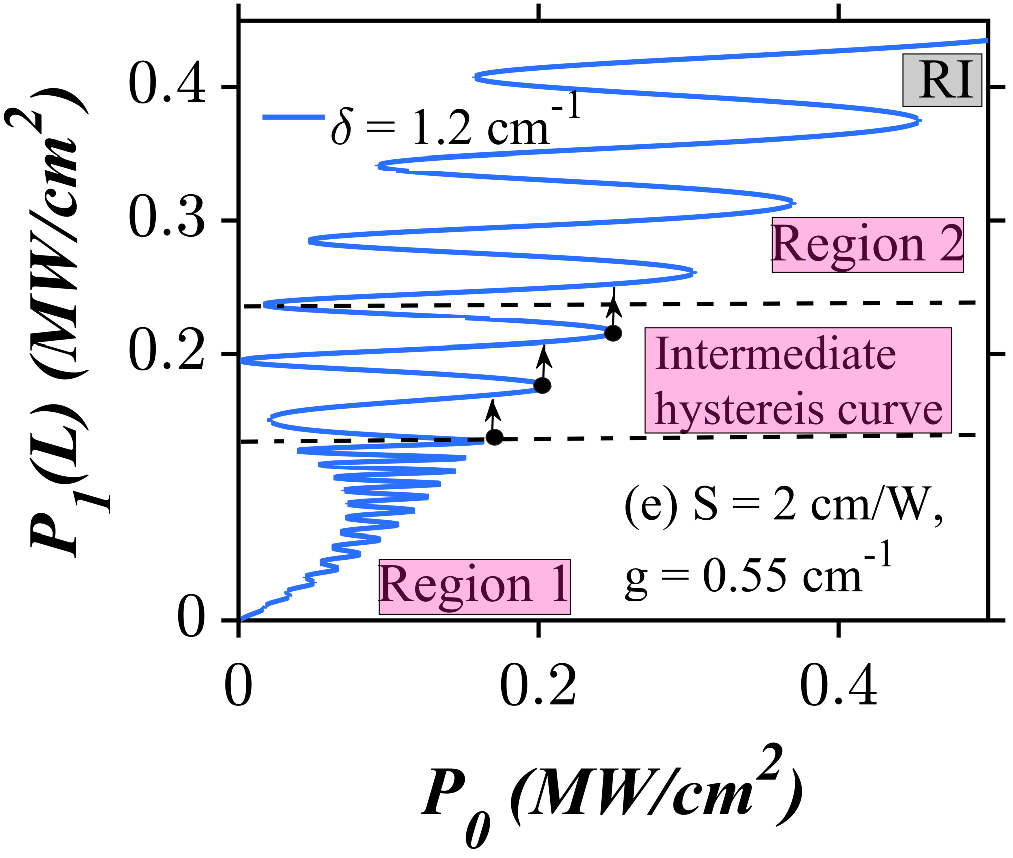}\includegraphics[width=0.25\linewidth]{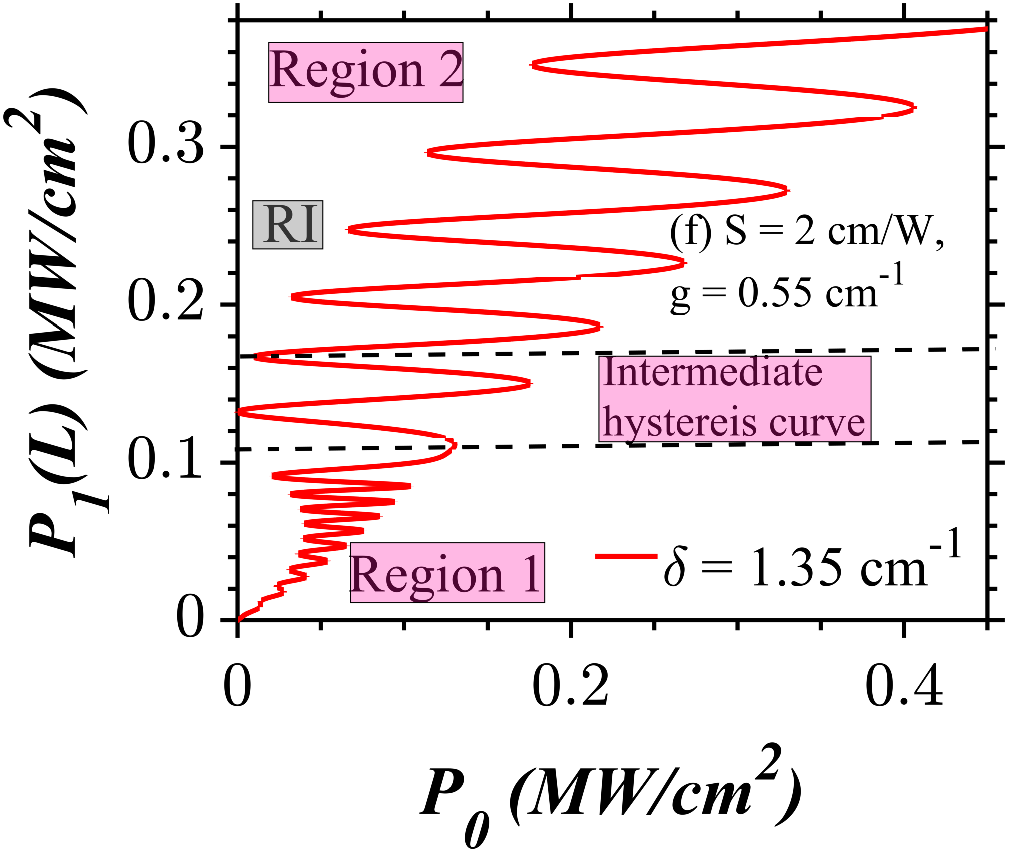}\includegraphics[width=0.25\linewidth]{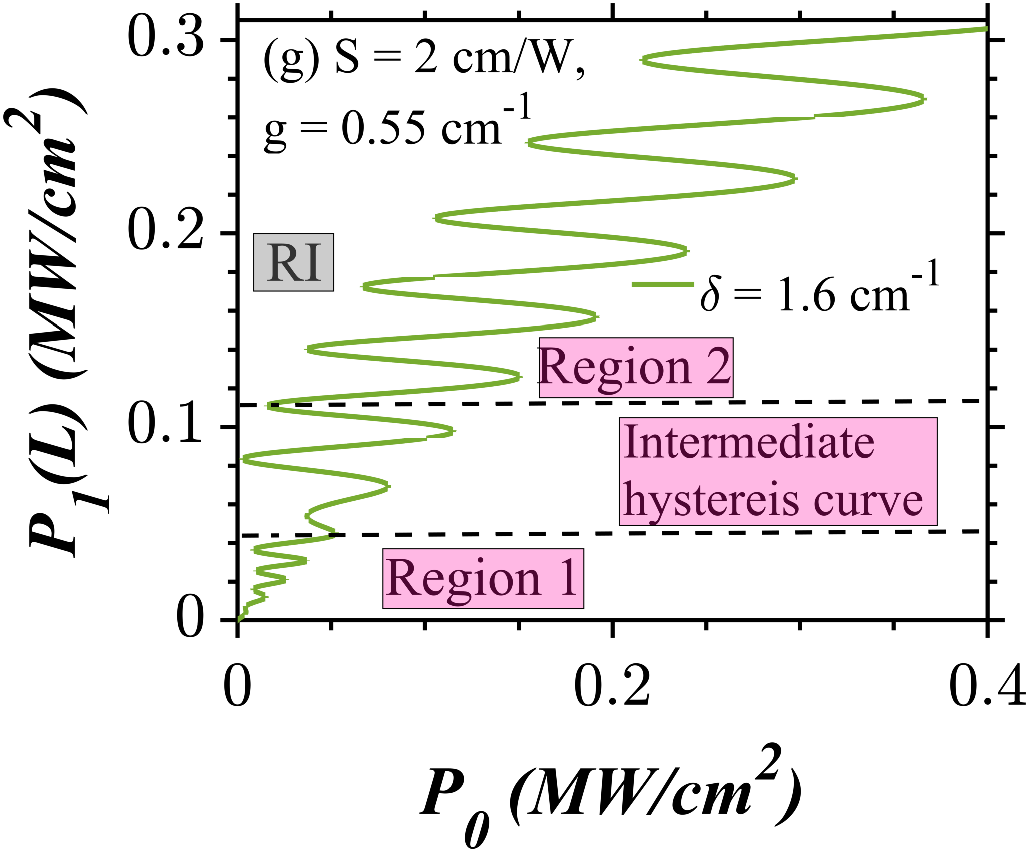}\includegraphics[width=0.25\linewidth]{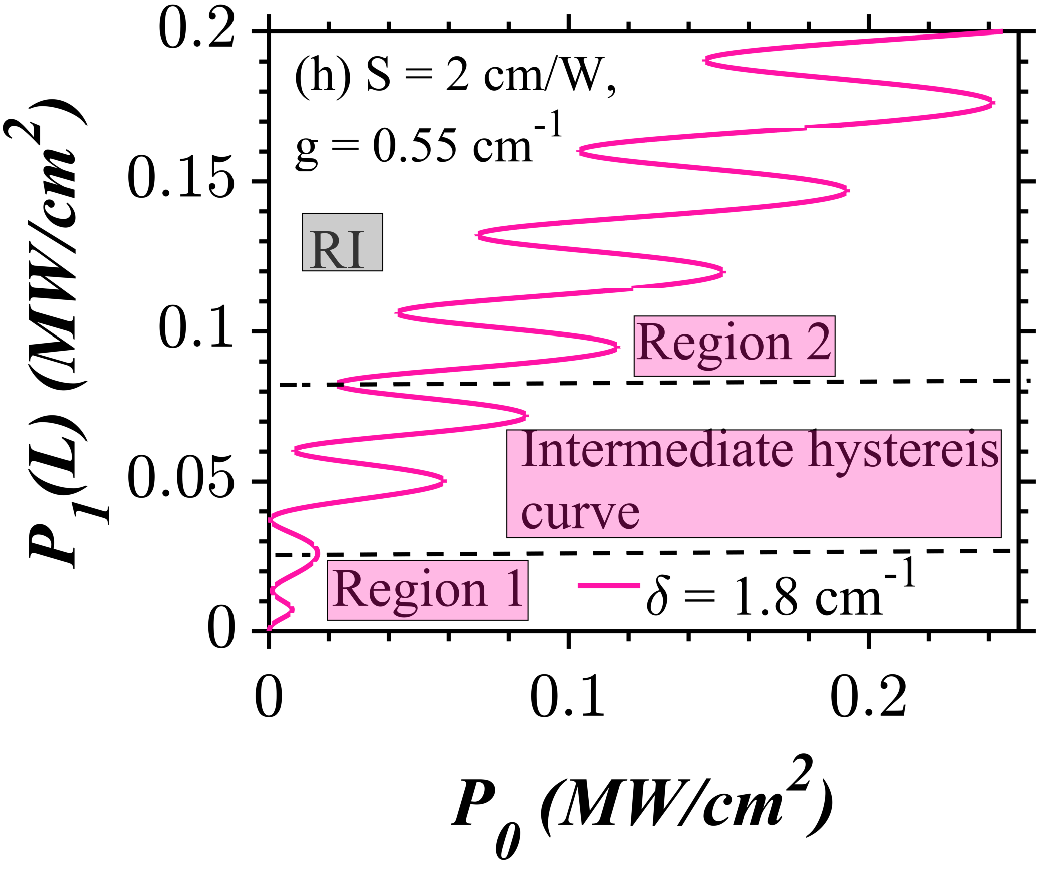}
 	\caption{Input-output characteristics of a broken PTFBG with SNL at $L = 70$ $cm$ and $S = 2$ $cm/W$. (a) Ramp-like OM, (b) mixed OM, (c) and (d) low-power ramp-like OM with vortex-like envelope,  (e), (f) low-power mixed OM with vortex-like envelope, (g) and (h) low-power mixed OM curves with more number of S-shaped and less ramp-like OM curves. The direction of light incidence is left in (a, b), and is right (c-f). }
 	\label{fig8}
 \end{figure*}

 Numerical investigations confirm that as the device length increases, the number of stable states of ramp-like and mixed OM curves increases, as shown in Figs. \ref{fig8}(a) and (b), respectively, under the left light incidence condition. Note that the frequency detuning aids in the transformation of ramp-like OM curves into mixed OM curves.
 
 Under the condition of right light incidence, the system displays ramp-like and mixed OM curves with a \emph{vortex-like envelope}, as depicted in the middle and bottom panels of Fig. \ref{fig8}. The switch-up and switch-down intensities of these OM curves drift towards higher and lower intensity sides, respectively, marking an unusual phenomenon. Typically, in any OM curve, they drift towards the higher-intensity side, which is not the case here. The simultaneous shift in the switch-up and switch-down intensities towards higher and lower intensity sides results in a vortex-like envelope, as shown in Figs. \ref{fig8}(c) -- (f). An intermediate hysteresis curve with a \emph{near-zero switch-down} intensity separates the ramp-like portion of a mixed OM curve in region 1 from the S-shaped OM curves in region 2. It exhibits a sharp increase in output intensities in the lower branch, contrasting with a gradual increase in the upper branch. Notably, the formation of OM curves, where one of its branches features near-zero switch-down intensities, has been reported in anti-directional couplers \cite{govindarajan2019nonlinear} and PTFBGs with inhomogeneous NL profiles \cite{sudhakar2022inhomogeneous}. However, the concurrent presence of mixed OM curves showcasing a vortex-like envelope and a hysteresis curve with near-zero switching intensities is a \emph{unique discovery}, not mentioned in earlier studies. This remarkable phenomenon is feasible due to the intricate interplay between SNL and the right light incidence condition. Increasing the value of detuning parameter decreases the switching intensities pertaining to different stable branches of these peculiar OM curves with a vortex-like envelope, as depicted in Fig. \ref{fig8}(f). Furthermore, there is a considerable reduction in the number of ramp-like stable branches in region 1, and the number of S-shaped hysteresis curves in region 2 increases for larger values of the detuning parameter, as shown in Figs. \ref{fig8}(g) and (h).
 
 \section{Dynamics at the unitary transmission point ($\kappa = g$)}
 \begin{figure}
 	\centering	\includegraphics[width=0.45\linewidth]{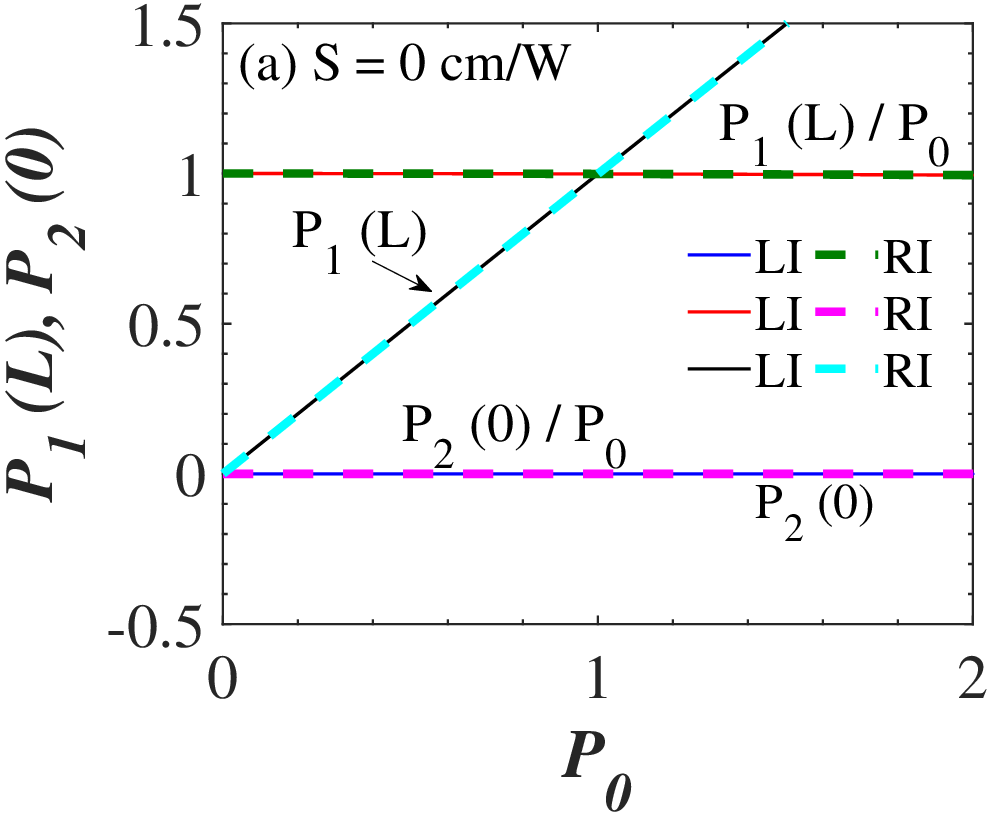}	\includegraphics[width=0.45\linewidth]{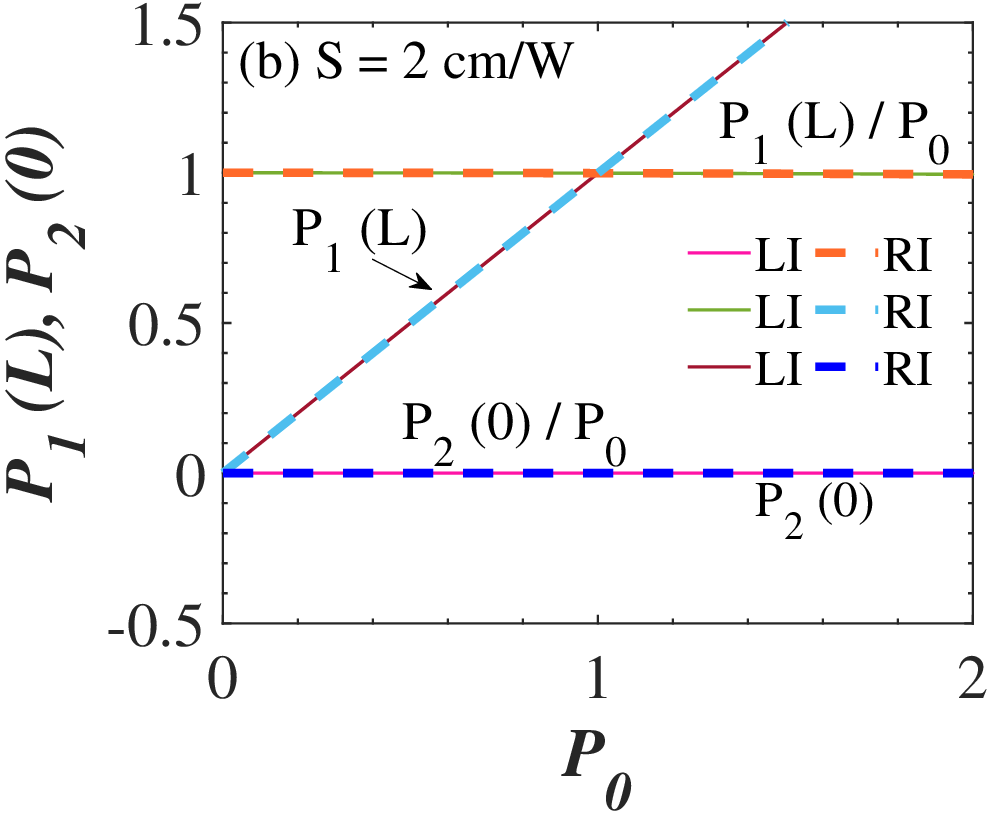}
 	\caption{Input-output characteristics of a broken PTFBG with SNL (a) $S = 0~ cm/W$ (linear) and (b) $S = 2~ cm/W$ (nonlinear) at $L = 20$ $cm$ and $\kappa = g = 0.4~ cm^{-1}$. The system behavior remains the same for all values of $\delta$ in both linear and nonlinear regimes.}
 	\label{fig10}
 \end{figure}
  From the numerical results depicted in Fig. \ref{fig10}(a), we observe that the transmission coefficient [$P_1(L)/P_0$] reaches unity and the   transmitted intensity [$P_1(L)$] shows a linear increase against the input power ($P_0$) at the $\mathcal{PT}$ -symmetric breaking point in the linear regime ($S = 0 \text{ cm/W}$) in a linear PTFBG for both left and right light incidences. Additionally, from the same figure, we observe that the reflected intensity ($P_2(0)$) and reflection coefficient ($P_2(0)/P_0$) are always zero. One may question whether the system will exhibit the same dynamics in the nonlinear regime in the presence of  SNL parameter. To confirm this, we set the SNL parameter to be $S = 2 \text{ cm/W}$ in the simulations. Figure \ref{fig10}(b) reveals that even with the inclusion of the SNL parameter ($S = 2 \text{ cm/W}$), the transmitted (reflected) intensity and transmission (reflection) coefficient remain the same as in the linear regime, similar to Fig. \ref{fig10}(a). This once again proves that the operation of the system at the unitary transmission point entirely depends on the balance between the coupling and gain-loss coefficients ($\kappa = g$) (which further affirms that the notion of $\mathcal{PT}$ -symmetry is specifically limited to the linear case alone) and is independent of any variations in other system parameters, including SNL.

	\begin{table}[t]
	\caption {Types of hysteresis curves admitted by the PTFBG with SNL. Note that the output parameters $P_{th}^{\uparrow}$, $P_{th}^{\downarrow}$ and $\Delta^{\uparrow\downarrow}$ decrease and increase, with an increase in the detuning ($\delta$) and SNL ($S$) parameters, respectively.  }
	\begin{center}
			\begin{tabular}{c c c}
				\hline
				\hline
				{Nature of the}&{Classification}&{Nature of variations} \\
				{hysteresis}&{of the}&{in properties of} \\
				{curve}&{curve}&{the OM curve}\\
				
				\hline

				
				{ramp-like }&{Type -- I}&{$P_0$ vs $P_1(L)$: sharp}\\
				{OB and}&{}&{increase} \\
				{OM curves}&{}&{$\Delta^{\uparrow\downarrow}$ vs $P_0$: increases} \\
				
				
				{[Figs. \ref{fig1}, \ref{fig2}, \ref{fig6}, \ref{fig8}(a)]}&{}& {}\\
					{}&{}& {}\\
				
				{mixed OM}&{Type -- II}&{ $P_1(L)$ vs $P_0$: sharp}\\
				{curves  }&{}&{at low and}\\
				{[Figs. \ref{fig3}, \ref{fig7}, \ref{fig8}(b), \ref{fig8}(g) and (h)]}&{}&{gradual increase at }\\
				{}&{}&{high intensities}\\
				
				{}&{}&{$\Delta^{\uparrow\downarrow}$ vs $P_0$:increases}\\
				{}&{}&{ at low and }\\
				{ }&{}&{and decreases at high}\\
				{ }&{}&{intensities}\\
				{}&{}&{}\\
				
					{S-shaped}&{Type -- III}&{$P_0$ vs $P_1(L)$: gradual}\\
				{OB curves }&{ }&{increase}\\
				{[Figs. \ref{fig4} and \ref{fig5}]}&{}&{$\Delta^{\uparrow\downarrow}$ vs $P_0$: increases}\\
				{}&{}&{ }\\
				
				{ramp-like OM curves}&{Type -- IV}&{same as Type - I}\\
				{with vortex-like  }&{}&{}\\
				{envelope [Figs. \ref{fig8}(c) and (d)]}&{}&{}\\
				{}&{}&{}\\

				{mixed OM curves}&{Type -- V}&{same as Type - II} \\
				{with vortex-like  }&{}&{}\\
				{envelope [Figs. \ref{fig8}(e) and (f)]}&{}&{}\\
				{}&{}&{}\\
				
				\hline\hline
		\end{tabular}
		\label{tab2}
	\end{center}
\end{table}

\begin{table}[htbp]
	\caption{Do Type - I to V curves exist in different $\mathcal{PT}$-symmetric operating regimes in PTFBGs without \cite{raja2019multifaceted,PhysRevA.100.053806} and with SNL?. Regimes I, II, and III represent Conventional ($g = 0$), unbroken ($\kappa > g$), and  broken ($\kappa < g$) regimes, respectively.}
	\centering
	\begin{tabular}{c c c c c c c}
		\hline \hline
		\multirow{2}{*}{} & \multicolumn{3}{c}{PTFBGs without SNL} & \multicolumn{3}{c}{Present System} \\ \cline{2-7} 
		& Regime I & II &III & Regime I & II & III \\ 
		Type - I &No  &No  &Yes  &Yes  &Yes  & Yes \\ 
		Type - II &No &No &No &No  &Yes  &Yes \\ 
		Type - III &Yes &Yes &No  &No &Yes &Yes \\ 
		Type IV	and V &No &No &No  &No  &No & Yes \\ \hline
	\end{tabular}
	
	\label{tab3}
\end{table}

	 	\begin{table}[h]
	 		\caption{Role of different control parameters on the critical switch up ($P_{th}^{\uparrow}$), down ($P_{th}^{\downarrow}$) intensities, and hysteresis width ($\Delta^{\uparrow\downarrow}$) of the OM curves shown by PTFBGs with SNL in the unbroken regime.}
	 		\begin{center}
	 			\begin{tabular}{c c c c}
	 				\hline
	 				\hline
	 				{Type of}&{Increase in}& {Impact on} & {Impact on}  \\
	 					{OB/OM }&{the control }&{the switching }&{the hysteresis}\\
	 						{curve}&{parameter}&{intensities}&{ width}\\
	 				\hline

	 				{Type -- I}&{S}&{increases}& {increases} \\

	 				{Type -- I}&{g}&{increases}& {increases} \\

	 				{Type -- II}&{g and S}&{increases}&{increases}\\
	 				
	 				{Type -- II}&{$\delta$}&{decreases}&{decreases}\\

	 				\hline\hline
	 			\end{tabular}
	 			\label{tab4}
	 		\end{center}
	 	\end{table}
 	
 	\begin{table}
 		\caption{Hysteresis curves in the unbroken and broken $\mathcal{PT}$- symmetric regimes generated by PTFBGs with SNL. Note that one can adjust the value of $S$ from 0.5 to 2 $cm/W$, categorized as low ($S < 0.5$), moderately high ($0.5 < S < 1$), high ($1 < S < 1.5$), and very high ($S > 1.5$ $cm/W$). Similarly, values of the detuning parameter in the  range $0.5 < \delta < 0.75$ $cm^{-1}$ are considered moderately high, while those below and above are labeled as low and high, respectively.}
 		\begin{center}
 		
 				\begin{tabular}{c c c}
 					\hline
 					\hline
 					{Classification of}&{Unbroken $\mathcal{PT}$} & {Broken $\mathcal{PT}$}  \\
 					{the curve}&{ regime } & {regime}  \\
 					\hline
 					{Type - I}&{values of $S$: all}&{low}\\
 					{}&{values of $\delta$: high}&{ low}\\
 					{}&{range of $\delta$: narrow }&{ narrow}\\
 					{}&{}&{}\\
 					{}&{}&{}\\
 					{Type - II}&{values of $S$: all}&{moderately high}\\
 					{}&{only at $\delta = 0$ }&{ high}\\
 					{}&{}&{range of $\delta$: broad}\\
 					{}&{}&{}\\
 					
 					{Type -- III}&{values of $S$: all }&{high and very high}\\
 					{}&{values of $\delta$: low}&{high}\\
 					{}&{range of $\delta$: narrow }&{broad}\\
 					{}&{}&{}\\
 					
 					{Type -- IV}&{cannot occur}&{values of $S$: very high }\\
 					{}&{}&{values of $\delta$: $<1$ $cm^{-1}$}\\
 					{}&{}&{under RI}\\
 					
 					{}&{}&{}\\
 					
 					{Type -- V}&{cannot occur}&{ values of $S$: very high }\\
 					{}&{}&{values of $\delta$: $>$1  $cm^{-1}$}\\
 					{}&{}&{under RI}\\
 					{}&{}&{}\\

 					\hline\hline
 			\end{tabular}
 			\label{tab5}
 		\end{center}
 	\end{table}
 		\section{Conclusions}
 	\label{Sec:6}
	Tables \ref{tab2} -- \ref{tab5} summarize all the outcomes of the present work. The system under study with saturable nonlinearity has exhibited quite a few unusual and unique optical bi- and multi- stable states. For instance,  instead of admitting an S-shaped (ramp-like) hysteresis curve, conventional and unbroken (broken) FBGs have remarkably revealed ramp-like (S-shaped) OB and OM curves.   The system admitted mixed OM curves in which ramp-like first stable states preceded the S-shaped stable states for the values of detuning parameter closer to the Bragg wavelength. In both operating regimes, the system exhibits various types of OB and OM curves, including S-shaped, ramp-like, and mixed OM curves. In the conventional PTFBG systems without SNL, it is not feasible to generate ramp-like OB (OM) curve in the unbroken $\mathcal{PT}$ -symmetric regime and S-shaped OB (OM) in the broken $\mathcal{PT}$ -symmetric regime. While the overall behavior appears similar in both the regimes, the required input power to generate these curves differs. Specifically, S-shaped OB and OM curves form at lower input powers in the unbroken regime compared to the broken regime. Dramatic reduction in the switch-up and down intensities of the various OB (OM) curves occurred under a reversal in light incidence direction. In particular, they fell below $1.1$ $kW/cm^2$ in the case of an S-shape OB curve with higher values of detuning, NL, and gain-loss parameters, which must be the lowest-switching intensities in the context of PTFBGs. This distinction applies to the other OB and OM curve types discussed in the manuscript as well. The range of detuning and SNL parameters for the occurrence of various curves differs between the unbroken and broken regimes, as detailed in Table \ref{tab4}. The transition between different OB (OM) curves (under the variation in the detuning parameter) also varies between these regimes, highlighting distinct bistable responses. In the unbroken $\mathcal{PT}$ -symmetric regime, the OB curves transition from ramp-like to S-shaped as the detuning parameter increases from zero (synchronous wavelength) to high values. For detuning values in between, the system reveals mixed OB and OM curves. Conversely, in the broken $\mathcal{PT}$ -symmetric regime, the system transitions from the S-shaped curves at lower detuning values to the ramp-like curves at intermediate values and mixed OB curves at higher detuning values.
	
	  The numerical results presented here confirm that the presence of SNL in a PTFBG opens a road map to control light with light in diverse fashions. Also, they indicate that the PTFBGs are not only interesting from a theoretical perspective, as they appear to be attractive platforms for the practical realization of ultra-low power switching, thanks to the number of independent approaches they offer to control the switching intensities. 
	 
	 The concept of right light incidence enabled the realization of different types of OM curves at low intensities. In this light incidence condition, the switch-up and down intensities corresponding to different stable branches of a ramp-like OM and mixed OM curves demonstrated a drift towards higher and lower intensity sides, respectively, leading to a novel vortex-like envelope. Such a drift stimulated a near-zero switch-down phenomenon in one of the intermediate hysteresis curves. To sum up, the PTFBG with SNL offers numerous possibilities to control light with light. 
	 
	 In the continuously evolving field of fiber optic transmission systems, there is an increasing demand for processing multiple packets of information with a single, compact device offering high tunability and reconfigurability \cite{hill2004fast,adams1987physics,alexoudi2020optical,adams1987physics}. The advancements in all-optical memories, especially those adopting multi-level configurations, have demonstrated notable progress over the last decade \cite{tohari2020optical,zhang2017optical,daneshfar2017switching,temnykh2010optical}. The features observed in the OM curves within the proposed system suggest an alternative route for fabricating optical memories, potentially enabling high information density in low-power applications. Mixed OM curves, with gradual transitions between ramp-like and S-shaped portions of the hysteresis curve, may facilitate more controlled switching between memory states, reducing errors in data transmission and processing during read and write operations. The dynamics of ramp-like and mixed OM curves, with an unusual vortex-like envelope, can lead to unconventional memory functionalities, inspiring innovative designs in the near future. This complexity, easily tailored by adjusting system parameters, offers the potential for dynamic and re-configurable memory systems, meeting specific requirements in optical computing applications.

	\section*{Acknowledgements}
	SVR is supported by the  Department of Science and Technology (DST)-Science and Engineering Research Board (SERB), Government of India, through a National Postdoctoral Fellowship (Grant
	No. PDF/2021/004579). AG acknowledges the  support from the University Grants Commission (UGC), Government of India, for providing a Dr. D. S. Kothari Postdoctoral Fellowship (Grant
	No. F.4-2/2006 (BSR)/PH/19-20/0025). ML wishes to thank the DST-SERB for the award of a DST- SERB National Science Chair in which AG is now a Visiting Scientist (Grant No. NSC/2020/000029).

\end{document}